\documentclass[%
aps,twocolumn,floatfix,showpacs,superscriptaddress,amsmath,amssymb]{revtex4-1}
\usepackage{graphicx}
\usepackage{dcolumn}
\usepackage{bm}
\graphicspath{ {./images/} }
\usepackage{color}
\usepackage{hyperref}
\usepackage{footnote}
\usepackage{wrapfig}
\usepackage{subcaption}
\usepackage[T1]{fontenc}

\usepackage{amsmath}
\usepackage{amssymb}
\usepackage{amsthm}
\usepackage{amsfonts}
\usepackage{braket}
\usepackage{tabularx} 

\begin{document}

\title{Data-driven design of a new class of rare-earth free permanent magnets}

\author{Alena Vishina}
\altaffiliation{Department of Physics and Astronomy, Uppsala University, Box 516, SE-75120, Uppsala, Sweden}
\email{alena.vishina@physics.uu.se}
\author{Daniel Hedlund}
\affiliation{Department of Materials Science and Engineering, Uppsala University, Box 35, 751 03 Uppsala, Sweden}
\author{Vitalii Shtender}
\affiliation{Department of Chemistry - {\AA}ngstr{\"o}m, Uppsala University, Box 538, 751 21, Uppsala, Sweden}
 \author{Erna K. Delczeg-Czirjak}
\affiliation{Department of Physics and Astronomy, Uppsala University, Box 516, SE-75120, Uppsala, Sweden}
\author{Simon R. Larsen}
\affiliation{Department of Chemistry - {\AA}ngstr{\"o}m, Uppsala University, Box 538, 751 21, Uppsala, Sweden}
\author{Olga Yu. Vekilova}
\affiliation{Department of Materials and Environmental Chemistry, Stockholm University, 10691 Stockholm, Sweden}
\affiliation{Department of Physics and Astronomy, Uppsala University, Box 516, SE-75120, Uppsala, Sweden}
\author{Shuo Huang}
\affiliation{Department of Physics and Astronomy, Uppsala University, Box 516, SE-75120, Uppsala, Sweden}
\author{Levente Vitos}
\affiliation{Applied Materials Physics, Department of Materials Science and Engineering, Royal Institute of Technology, Stockholm, SE-100 44, Sweden}
\affiliation{Department of Physics and Astronomy, Uppsala University, Box 516, SE-75120, Uppsala, Sweden}
\author{Peter Svedlindh}
\affiliation{Department of Materials Science and Engineering, Uppsala University, Box 35, 751 03 Uppsala, Sweden}
\author{Martin Sahlberg}
\affiliation{Department of Chemistry - {\AA}ngstr{\"o}m, Uppsala University, Box 538, 751 21, Uppsala, Sweden}
\author{Olle Eriksson}
\affiliation{Department of Physics and Astronomy, Uppsala University, Box 516, SE-75120, Uppsala, Sweden}
\affiliation{School of Science and Technology, {\"O}rebro University, SE-701 82 {\"O}rebro, Sweden}
\author{Heike C. Herper}
\affiliation{Department of Physics and Astronomy, Uppsala University, Box 516, SE-75120, Uppsala, Sweden}

\date{\today}
\begin{abstract}

A new class of rare-earth-free permanent magnets is proposed. The parent compound of this class is Co$_3$Mn$_2$Ge, and its discovery is the result of first principles theory combined with experimental synthesis and characterisation. The theory is based on a high-throughput/data-mining search among materials listed in the ICSD database. From ab-initio theory of the defect free material it is predicted that the saturation magnetization is 1.71 T, the uniaxial magnetocrystalline anisotropy is 1.44 MJ/m$^3$, and the Curie temperature is 700 K.
Co$_3$Mn$_2$Ge samples were  then synthesized and characterised with respect to structure and magnetism. The crystal structure was found to be the MgZn$_2$-type, with partial disorder of Co and Ge on the crystallographic lattice sites. From magnetization measurements a saturation polarization of 0.86 T at 10 K was detected, together with a uniaxial magnetocrystalline anisotropy constant of 1.18 MJ/m$^3$, and the Curie temperature of $T_{\rm C}$ = 359 K. These magnetic properties make Co$_3$Mn$_2$Ge a very promising material as a rare-earth free permanent magnet, and since we can demonstrate that magnetism depends critically on the amount of disorder of the Co and Ge atoms, a further improvement of the magnetism is possible. 
From the theoretical works, a substitution of Ge by neighboring elements suggest two other promising materials - Co$_3$Mn$_2$Al and Co$_3$Mn$_2$Ga. We demonstrate here that the class of compounds based on $T_3$Mn$_2$X (T = Co or alloys between Fe and Ni; X=Ge, Al or Ga) in the MgZn$_2$ structure type, form a new class of rare-earth free permanent magnets with very promising performance.

\end{abstract}
\maketitle\maketitle

\section{Introduction}

Rare earth (RE) permanent magnets dominate the market where high-performance magnetic materials are needed -  areas such as green-energy generation, including wind and wave power, electric vehicle motors and generators, and many more. 
At the same time, the heavier rare-earth elements which are necessary for obtaining the good magnetic characteristics of these materials (such as Pr, Nd, Sm, Tb, or Dy) \cite{perm_mag}, can only be mined with methods that leave an environmental footprint, are quite expensive, and are rapidly decreasing in availability. As a consequence many rare-earth elements are labelled critical.
In order to pursue green technologies that rely on high-performance magnets, there is a growing interest in finding new magnetic materials containing cheaper and less critical elements, while maintaining a similar high-performance shown by their RE counterparts.


In the last decades, many attempts have been made to design both RE-lean \cite{CeFeTi,FeCo} and RE-free magnetic materials.
The later include high magnetocrystalline anisotropy alloys (MnBi \cite{GABAY2020165860,KIM20171245} and MnAl \cite{FANG2016300,SHAFEIE2019229}), nanostructures \cite{mat1,mat3,mat5,mat7}, thin films \cite{film1,mat6}, alnico permanent magnets, and many more \cite{mat2,mat4,mat8,CUI2018118}. 
Techniques used for this search include both filtering through a large number of systems \cite{ours1,mach-learn,GAO2020355,FAYYAZI2017434} and investigating a single or a group of similar compounds \cite{mat2,mat4,mat9,mat10,mat11,mat12,mat13,mat14,mat15,PhysRevB.101.014426}.
In Ref. \cite{ours1} it was shown that a high-throughput density functional theory (DFT) approach can be effective in filtering through a large number of compounds in the search for new high-performance RE free permanent magnets, which would be time-consuming and  expensive to synthesize experimentally. Unfortunately, a commercially relevant class of rare-earth free compounds was not identified in Ref. \cite{ours1}.
In the current investigation, that data-filtering approach described in Ref. \cite{ours1} was broadened to involve a much larger starting set of compounds.

From a combination of data-mining efforts, using ab-initio electronic structure theory, materials synthesis, structural and magnetic characterization, we can report on a new class of rare-earth free permanent magnets, with very promising properties. In particular, we have identified Co$_3$Mn$_2$Ge, which is a new material that previously has not been considered as a permanent magnet.

\section{High-throughput and data-mining search}\label{sec:high}

An initial attempt to use high-throughput searches for RE-free permanent magnets was published in Ref.\cite{ours1}, employing a smaller subset of possible magnetic compounds. As described in this investigation, 
a high-performance permanent magnet must have ferromagnetic ordering, high saturation magnetization ($M_S> 1$ T), high Curie temperature ($T_{\bf C} >$ 400 K), a  large uniaxial MAE ($> 1$ MJ/m$^3$), and a magnetic hardness larger than 1 (fo details see the {\it Methods} section).
In the initial attempt of using high-throughput/data-mining searches, focus was put on materials with a stoichiometry necessarily containing one 3{\it d} and one 5{\it d} element \cite{ours1}. From a pure magnetic viewpoint, this study resulted in several promising materials. Unfortunately, none of the identified compounds in Ref.\cite{ours1} could serve as practical replacements for the known rare-earth based compounds, since they all contained expensive elements like Pt. 

In the present study, we have made an extended search to involve materials with a stoichiometry that contain two types of 3d-elements, without any restriction on the total number of elements in the system.
About a thousand materials from the ICSD database \cite{ICSD} contain two different 3d-elements in the chemical formula.
After the first high-throughput calculations of the electronic structure and magnetic moments, about 45 systems were identified to have a magnetic saturation field larger than a value $\sim$ 0.4 T. In the next step, all the materials with cubic symmetry were set aside, since spin-orbit effects are known to influence the MAE most effectively for non-cubic materials \cite{PhysRevB.39.865}.
We also performed an extensive literature search, to identify if any of the compounds identified in the intitial screening steps, had previously been reported as permanent magnet (e.g. the L1$_0$ phase of FeNi) or if the systems that were identified, are known to not be ferromagnetic. These compounds were also removed from the list of potential new rare-earth free permanent magnets.
For the remaining materials, the MAE was calculated.
Compounds with  planar or low magnetic anisotropy (MAE $< \sim$ 0.6 MJ/m$^3$) were not considered further, 
leaving only five materials whose saturation magnetization and MAE are high enough to be considered as high-performance permanent magnets. 

Note that the search criteria used here are somewhat less strict than the criteria of magnetic properties specified from practical aspects of high performance permanent magnets. The reason for using less strict criteria, is to not miss new classes of compounds, that have the potential to become technologically relevant, after e.g alloying or structural refinements. Using these criteria we have identified the following compounds as potential new permanent magnets: ScFe$_4$P$_2$, Co$_3$Mn$_2$Ge, Mn$_3$V$_2$Si$_3$, ScMnSi, and Cu$_2$Fe$_4$S$_7$ and we report in Table \ref{table:1} their calculated saturation magnetization, MAE, and magnetic hardness. All data were obtained from calculations at $T = 0$~K and for a ferromagnetic configuration. For this final list of materials, we also calculated the Heisenberg exchange interaction, and from
Monte Carlo simulations we identified the ground state magnetic configuration and, for materials with ferromagnetic configurations, the Curie temperature. Out of the five materials listed above, Co$_3$Mn$_2$Ge is the only one that is FM and has a $T_{\rm C}$ of 700 K. The rest of the materials were found to have a non-collinear magnetic structure and were for this reason not considered further. Some of the materials, discarded during the search, can be found in the {\it Appendix}, Table~\ref{table:4}. We'd like to note, that Co$_3$Mn$_2$Ge does have its magnetic hardness lower than 1. However, this parameter can be later on adjusted by alloying or other methods.
 
\begin{table}[h!]
\caption{Calculated MAE, saturation magnetization, Curie temperature, and magnetic hardness for the materials that can be considered as candidates for RE-free permanent magnets. NC stands for non-collinear spin structure  obtained from Monte Carlo simulations.}
\begin{tabular}{l l l l l l l} 
 \hline \hline
 Material & ICSD & Space & MAE & Sat. & $T_{\bf C}$  & $\kappa$ \\ 
  & number & group &  MJ/m$^3$ & magn., T & K & \\
 \hline
ScFe$_4$P$_2$ & 68525 & 136 & 0.71 & 0.65 & NC & 1.45 \\ 
Co$_3$Mn$_2$Ge & 52972 & 194 & 1.44 & 1.71 & 700 & 0.79 \\ 
Mn$_3$V$_2$Si$_3$ & 643689 & 193 & 0.85 & 0.73 & NC & 1.42 \\
ScMnSi & 86369 & 189 & 0.69 & 0.52 & NC & 1.79 \\
Cu$_2$Fe$_4$S$_7$ & 15973 & 51 & 0.90 & 0.37 & NC & 2.87 \\
 \hline \hline
\end{tabular}
\label{table:1}
\end{table}


\section{Experimental results}\label{sec:exp}

\subsection{Synthesis and structural characterisation}

\begin{figure}[h!]
 \centering
 \includegraphics[scale=0.30]{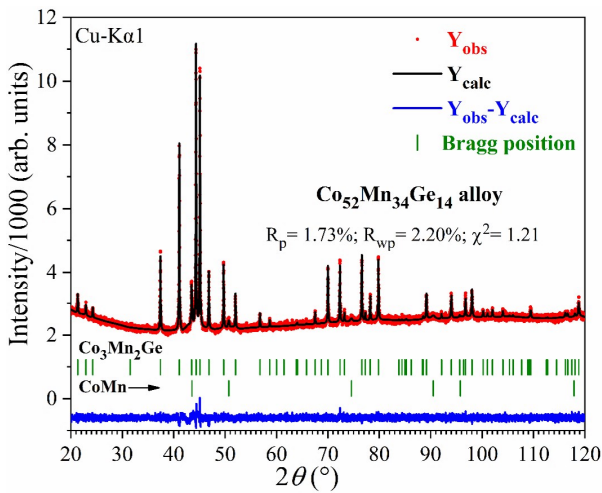}
 \caption{(Color online) Refined powder diffraction data of the synthesized Co$_{52}$Mn$_{34}$Ge$_{14}$ alloy. Observed (Y$_{obs}$), calculated
(Y$_{calc}$), difference (Y$_{obs}$-Y$_{calc}$) diffraction proﬁles and Bragg’s peaks positions for Co$_{3}$Mn$_{2}$Ge (95.4 wt.\%) and CoMn (4.6 wt.\%) phases are shown.}
 \label{Co3Mn2Ge_ref}
\end{figure}

Following the theoretical results presented above, focus was then put on synthesis of Co$_3$Mn$_2$Ge. Initial trials revealed the magnetic Heusler phase Co$_2$MnGe \cite{Co2MnGe_mag} as being the main competing phase in that region of the phase diagram. Preliminary structural refinement of the hexagonal Co$_3$Mn$_2$Ge phase (MgZn$_2$-type) in a multi-phase sample indicated disordering on the 6\textit{h} and 2\textit{a} sites implying intermixing between Co and Ge. EDS analysis on the multi-phase samples revealed the composition of the desired hexagonal phase to be in a small homogeneity region spanning Co$_{52}$Mn$_{34}$Ge$_{14}$ and Co$_{53.7}$Mn$_{31.7}$Ge$_{14.6}$. Samples following these stoichiometries were then synthesized. The Co$_{52}$Mn$_{34}$Ge$_{14}$ sample consisted of 95.4 wt\% of the hexagonal phase and 4.6 wt\% of CoMn, a Pauli paramagnet \cite{CoMn_Paramagnet}. The diffractogram and accompanying structural refinement is shown in Fig.  \ref{Co3Mn2Ge_ref}. The structure of the compound (R$_{Bragg}$ = 4.20 \%) is in good agreement with previous studies \cite{CoMnGe} having similar unit cell dimensions (V = 154.60(1) $\mathrm{\mathring{A}}^3$ compared to 154.61 $\mathrm{\mathring{A}}^3$ \cite{CoMnGe}). The final refinement resulted in a composition of Co$_{3.39}$Mn$_2$Ge$_{0.61}$ with intermixing of Co and Ge on the 6\textit{h} and 2\textit{a} sites.

Results from XRPD refinements and microstructural EDS results indicate that the structure is disordered with a tendency to contain more Co. Kuz'ma et al. \cite{CoMnGe} proposed both the ordered Mg$_2$Cu$_3$Si-type and disordered MgZn$_2$-type structures from their XRPD results but a clear atomic distribution could not be established. To determine the atomic distribution for the theoretical calculations, precision single crystal studies were conducted. Three variants of the structure were considered during refinement and are presented in Table \ref{table:SCXRD Ref} - the completely ordered structure, Co-Ge intermixing, and Mn-Ge intermixing. The structure solution obtained with SHELXT-2014 
quickly provided an ordered model for Co${_3}$Mn$_2$Ge. However, the wR$_2$ value, the EDS results, and the difference Fourier map all suggest that this model should be rejected. The model of Mn-Ge intermixing showed a larger goodness-of-fit and the calculated composition was too far from the measured to be accepted. The final model of Co-Ge intermixing overall shows the best parameters and is in agreement with other results. All atoms were refined anisotropically. Detailed crystallographic results of the SCXRD refinements can be found in the {\it Appendix}, Table~\ref{table:refine}.

\begin{table*}[t]
\centering
\caption{Results of single crystal refinements of the Co$_{3+x}$Mn$_2$Ge$_{1-x}$ compound. The Co-Ge disordered model (marked in bold) describes the measured data best.}
\begin{tabular}{l l l l l } 
 \hline \hline
 Parameters & Ordered & \textbf{Co-Ge disordered} & Mn-Ge disordered \\ 
 \hline
Occupation 6{\it h} & Co & \textbf{0.84Co+0.16Ge} & Co\\
 4{\it f} & Mn & \textbf{Mn} & Mn\\
 2{\it a} & Ge & \textbf{0.74Co+0.26Ge} & 0.4Ge+0.6Mn\\
\hline
Calculated formula & Co$_3$Mn$_2$Ge & \textbf{Co$_{3.24}$Mn$_{2}$Ge$_{0.76}$} & Co$_{3}$Mn$_{2.6}$Ge$_{0.4}$\\
Calculated composition (at.\%) & Co$_{50}$Mn$_{33.33}$Ge$_{16.67}$ & \textbf{Co$_{54}$Mn$_{33.33}$Ge$_{12.67}$} & Co$_{50}$Mn$_{43.33}$Ge$_{6.67}$\\
\hline
Goodness-of-fit on F$^2$ & 1.214 & \textbf{1.170} & 1.353 \\
\hline
Final R indices (I $>$ 2s(I)) & R$_1$ = 0.0432 & \textbf{R$_1$ = 0.0100} & R$_1$ = 0.0149\\
    & wR$_2$ = 0.1188 & \textbf{wR$_2$ = 0.0264} & wR$_2$ = 0.0360\\
\hline
R indices (all data) & R$_1$ = 0.0439 & \textbf{R$_1$ = 0.0111} & R$_1$ = 0.0160\\
    & wR$_2$ = 0.1192 & \textbf{wR$_2$ = 0.0268} & wR$_2$ = 0.0364\\
    \hline
Highest difference peak & 2.277 & \textbf{0.639} & 0.693\\
Deepest hole & -5.961 & \textbf{-0.420} & -1.003\\
1-$\sigma$ level & 0.569 & \textbf{0.119} & 0.158\\
 \hline \hline
\end{tabular}
\label{table:SCXRD Ref}
\end{table*}

\subsection{Experimental magnetism}
\begin{figure}[h!]
 \centering
 \includegraphics[scale=0.35]{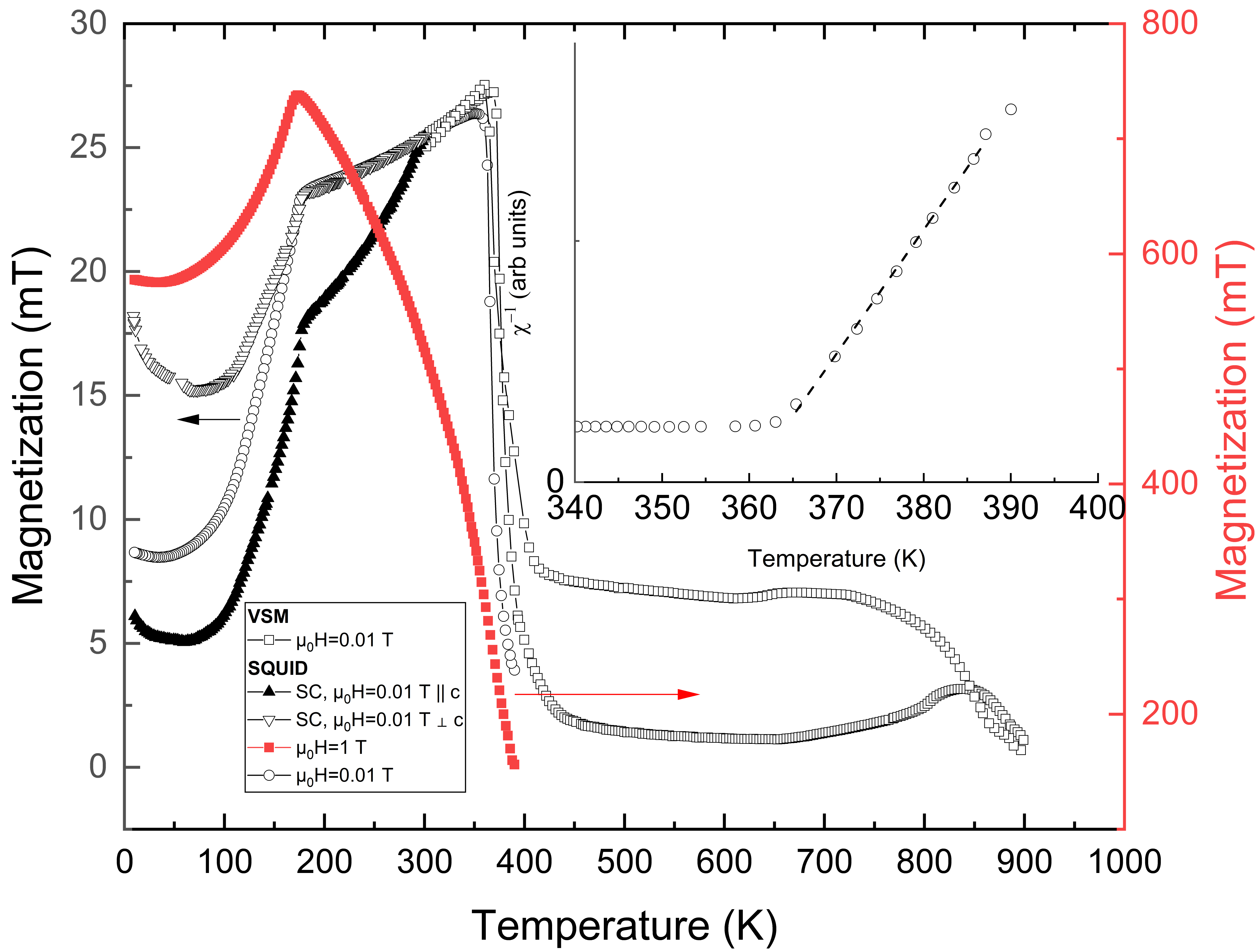}
 \caption{(Color online) Magnetization of Co$_3$Mn$_2$Ge as a function of temperature in applied magnetic fields of $\mu_0$H = 0.01 T (white open circles and white open rectangles) and $\mu_0$H = 1 T (red filled rectangles). The inset shows the Curie-Weiss fit for the inverse magnetic susceptibility with $T_{\rm C}=359$ K. The cusp in $\mu_0$H = 1 T and drop in magnetization around 175 K is attributed to a spin--reorientation ($T_{srt}$).}
 \label{M-T}
\end{figure}

The magnetization of the Co$_3$Mn$_2$Ge sample was measured as a function of temperature and magnetic field to determine the magnetic ordering temperature $T_{\rm C}$, the saturation magnetization, and the effective magnetic anisotropy constant of the material.
The temperature dependent magnetization measurements were performed at several applied magnetic fields and in the temperature region between 10 K and 900 K.
The results are shown in Fig.~\ref{M-T}. The magnetic ordering temperature, $T_{\rm C}=359$~K, was determined from the Curie-Weiss law fit to the temperature dependence of the inverse magnetic susceptibility (cf. inset in Fig.~\ref{M-T}).

\begin{figure}[h!]
 \centering
 \includegraphics[scale=0.49]{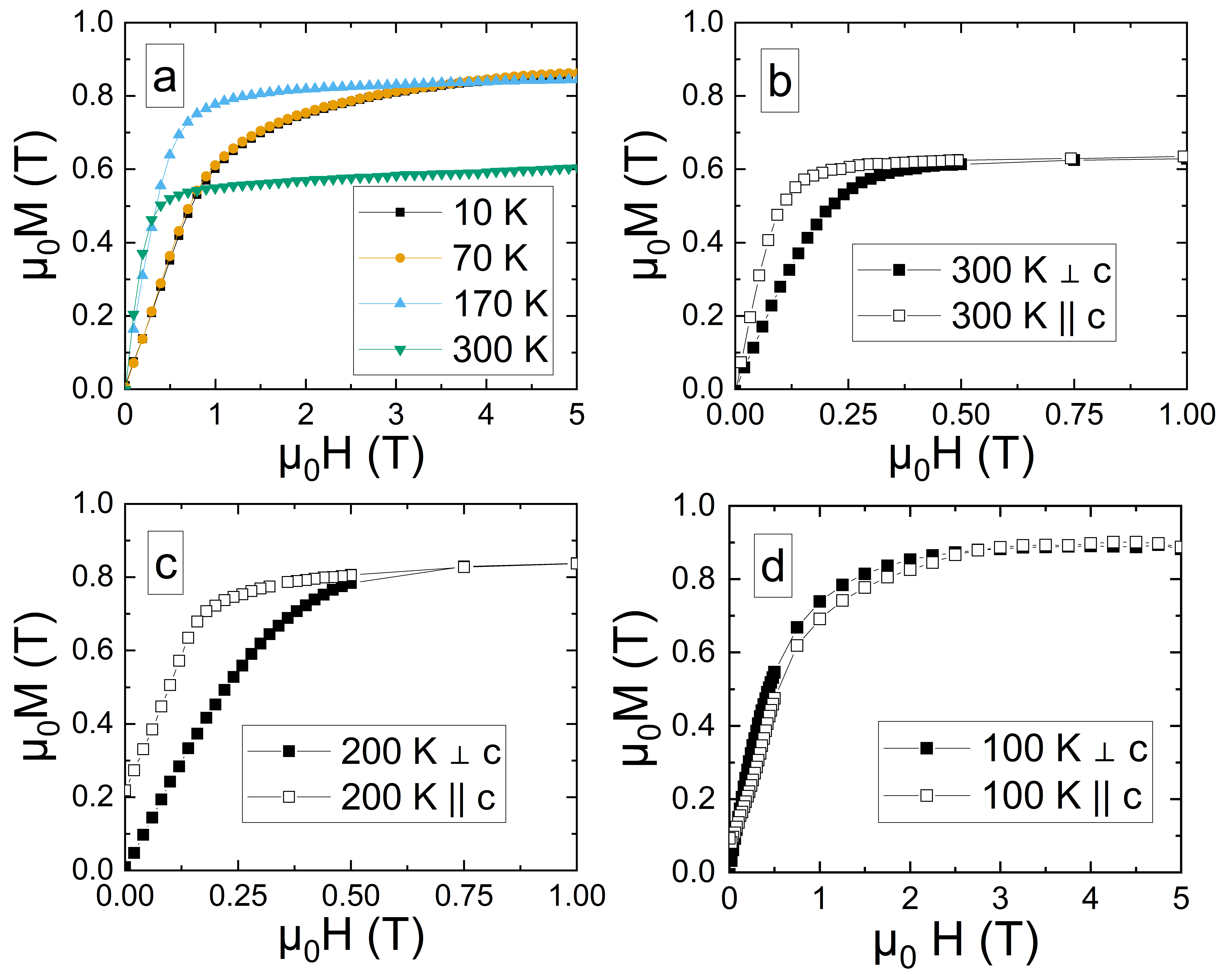}
 \caption{(Color online) (a) Magnetization of a bulk powder of Co$_3$Mn$_2$Ge, as a function of magnetic field at 10 K, 70 K, 170 K and 300 K. (b,c,d) Isothermal magnetization of single crystals of Co$_3$Mn$_2$Ge at 300 K, 200 K and 100 K. Filled symbols show measurements where the magnetic field is perpendicular ($\bot$) to the crystallographic {\it c}--axis whereas open symbols show measurements with the magnetic field parallell ($\parallel$)with the {\it c}--axis. }
 \label{M-H}
\end{figure}

The isothermal magnetization curves are shown in Figure~\ref{M-H}. Note that Figure~\ref{M-H}(a) shows the result for bulk powders whereas Figure~\ref{M-H}(b--d) corresponds to single crystal measurements at different temperatures along ($\parallel$)  the crystallographic c--axis and perpendicular ($\bot$) to it.
In Figure~\ref{M-H}(a) the isothermal magnetization for the powder sample, measured at 10 K and 70 K, shows the same approach to saturation and reach the same saturation polarization of 0.86 T.
The isothermal magnetization measured at 170 K reaches a saturation polarization close to that recorded at lower temperature; whereas the saturation polarization measured at 300 K reaches a value of 0.60 T.

\begin{figure}[h!]
    \centering
    \includegraphics[scale=0.35]{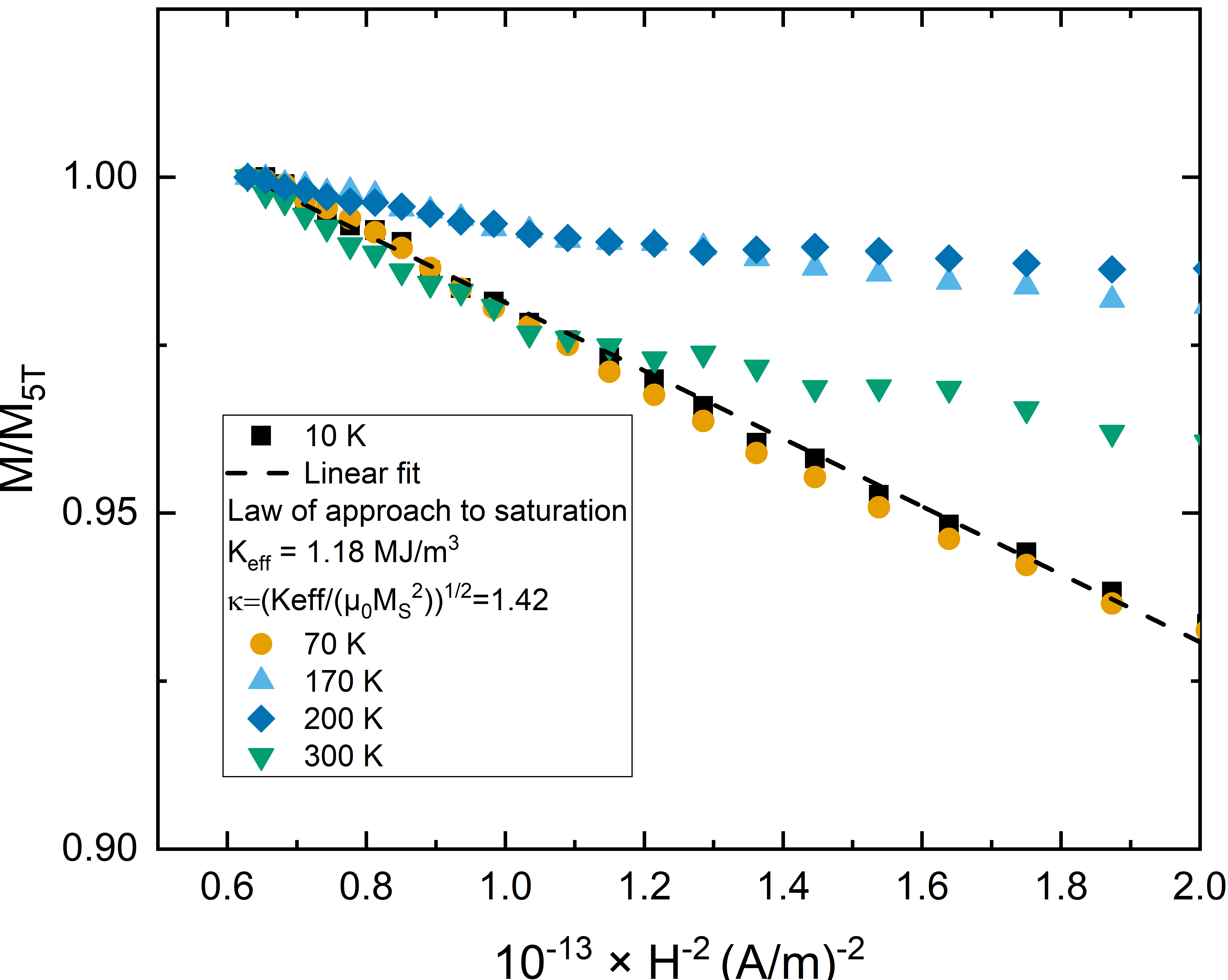}
    \caption{(Color online) Approach to saturation plotted as the magnetization normalized with the value measured at 5 T as a function of inverse applied field squared for the bulk sample of Co$_3$Mn$_2$Ge.}
    \label{LAS}
\end{figure}

To analyze the behavior of the magnetic anisotropy and to compare to the value calculated theoretically, we have used the law of approach to saturation to calculate the effective anisotropy constant at different temperatures. 
The results are shown in Figure \ref{LAS}. 
At 10 K and 70 K, the law of approach to saturation yields an effective anisotropy constant of 1.18 MJ/m$^3$ and the  magnetic hardness parameter, $\kappa$, becomes 1.42. The experimental results of magnetic configuration, saturation field, and magnetic anisotropy hence seem consistent with the theoretical predictions of this compound, which is gratifying. 

A closer inspection of the results presented in Fig.~\ref{M-T} reveal that below $\sim$ 175 K the magnetization drops considerably when the applied field is small (0.01 T), whereas a cusp--like feature is seen in an applied magnetic field of 1 T, both for single crystal material and bulk samples.
Such features can be attributed to a change in magnetocrystalline anisotropy with temperature and we interpret this to be a spin--reorientation temperature $T_{srt}$, from  easy--axis to easy--cone anisotropy, a feature which is not uncommon for permanent magnets (e.g. Fe$_5$SiB$_2$ \cite{Cedervall2016}, MnBi \cite{McGuire2014}, Cr$_{0.9}$B$_{0.1}$Te \cite{He2020}, Gd \cite{Corner1975} and the celebrated Nd$_2$Fe$_{14}$B \cite{Tokuhara1985}).  
Before presenting further results supporting a claim of a temperature induced spin--reorientation, we draw attention to the fact that for a hexagonal uniaxial material the magnetic anisotropy energy can be expressed as $MAE = K_1 \cos^2{\theta} + K_2 \sin^4{\theta}$, where  $V$ is the volume, $K_i$ are the anisotropy constants, and $\theta$ is the angle between the magnetization vector and the hexagonal c--axis \cite{krishnan2016fundamentals}. 
Depending on the values of $K_1$ and $K_2$ three cases can be distinguished; the material has an easy--axis, an easy--plane or an easy--cone magnetic anisotropy. 
The easy-cone state is the most favorable one when $-2K_2 \leq K_1 \leq 0$. 
The temperature dependence of the anisotropy constants ($K_i(T)$) may be strong at low temperature and is usually described by a power--law behavior, $K_i(T) \propto  M(T)^\alpha$ \cite{Akulov1936,Callen1966}, where $M(T)$ is the magnetization and $\alpha$ is a constant. The temperature dependence of the anisotropy constants can then lead to a spin--reorientation, e.g. from an easy--axis to an easy--plane or an easy--axis to an easy--cone anisotropy.

The material under investigation here shows characteristics of an easy--axis to easy--cone spin--reorientation around 175 K. We support this claim with the aid of isothermal magnetization curves recorded at temperatures below and above $T_{srt}$ on the single crystals and bulk powders.
The single crystal isotherms presented in Figure~\ref{M-H}(b--d) indicate that the easy--cone state is present at 100 K and below. 
Looking at the magnetization curves at 300 K (Figure~\ref{M-H}(b)) we see that it is easier to magnetize the crystal along the c--axis than perpendicular to it.
This shows that the material is magnetically uniaxial at 300 K. 
The same applies at 200 K (Figure~\ref{M-H}(c)), whereas at 100 K the measurements show that it is as easy to magnetize the material parallel or perpendicular to the crystallographic {\it c}--axis. 
However, it should be noted that at low field ($< 0.5$ T) there is a deviation from a linear magnetic response in measurements performed parallel or perpendicular to the crystallographic c--axis. 
The non--linear behaviour at low fields, together with the low field magnetization versus temperature measurements support the easy--cone state \cite{Fromter2008}. 

Figure~\ref{LAS} supports the conclusion of a more complicated behaviour for the magnetization of Co$_3$Mn$_2$Ge.
It should be noted, that in order to extract an anisotropy constant using the law of approach to saturation,  averaging over a random distribution of crystal directions yields the constant $\beta=4/15$ \cite{Herbst1998}. 
The effective anisotropy constants and magnetic hardness parameter should thus be seen as the tentative descriptions of the magnetocrystalline anisotropy at low temperatures.

Summarizing the experimental results so far 
we conclude that the theoretical prediction of Co$_3$Mn$_2$Ge as a potential rare-earth free permanent magnet is likely. The saturation magnetization, uniaxial magnetic anisotropy, and magnetic hardness parameter obtained from experiments are consistent with the theoretical predictions. It must be noted that a more detailed picture is present in the experiment, with a temperature stabilized easy-cone state. This state was not specified in the theoretical high-throughput screening search, and could therefore not be expected from the theoretical prediction. However, the easy axis, order of magnitude of the anisotropy, the saturation magnetization, and ordering temperature should agree, and here the theory has proven useful in finding a new material, with a potential to be used as a permanent magnet. We return to the easy cone state, below, with a theory that takes disorder into account, and show that calculations then reproduce experimental observations.



\section{Disordered C\lowercase{o}$_3$M\lowercase{n}$_2$G\lowercase{e}}\label{sec:disorder}

\begin{figure}[h]
\begin{subfigure}{0.5\textwidth}
\includegraphics[scale=0.42]{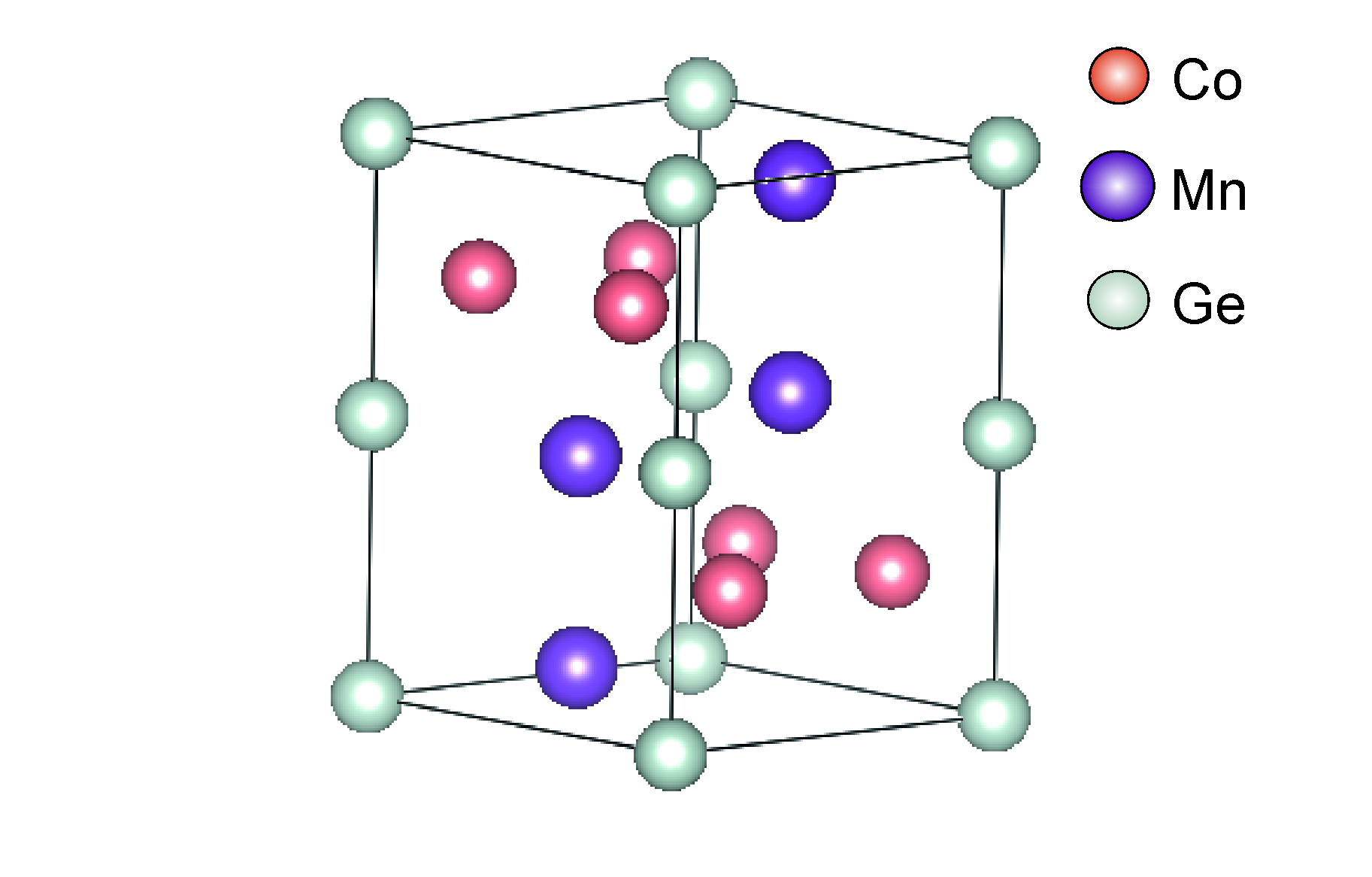} 
\label{order}
\end{subfigure}
\begin{subfigure}{0.5\textwidth}
\includegraphics[scale=0.37]{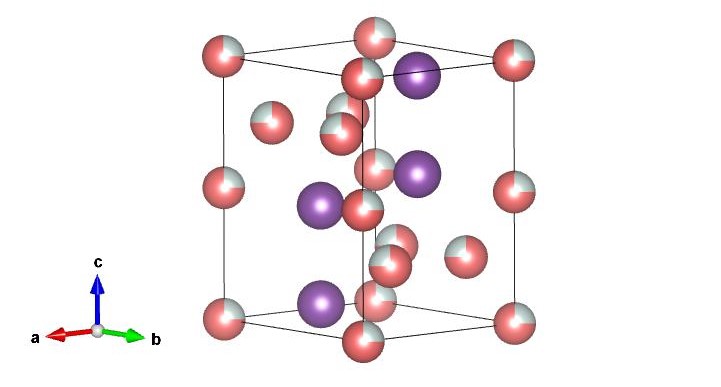}
\label{disorder}
\end{subfigure}
 \caption{(Color online) Ordered Co$_3$Mn$_2$Ge (as per ICSD) with Co atoms occupying 6{\it h} (light red balls), Mn atoms taking 4{\it f} (violet balls), and Ge (grey balls) occupying 2{\it a} atomic sites (top); and the disordered Co$_3$Mn$_2$Ge structure obtained experimentally with 50-50\% intermixing between the Co and Ge sites (bottom).}
\label{structure}
\end{figure}

According to the experimental reports, there may be a 50-50\% intermixing of Co and Ge atoms on 6{\it h} and 2{\it a} \cite{CoMnGe} Wyckoff positions (see Fig.~\ref{structure}), in addition to the excess of Co in comparison to Ge (see Table~\ref{table:SCXRD Ref}), in the experimental samples. This structure and composition differ from the one reported in the ICSD \cite{ICSD} database, where Co atoms are reported to occupy 6{\it h}, Mn atoms take 4{\it f}, and Ge occupies 2{\it a} atomic sites. It should be noted, that
disorder of the type detected in our experiments is often observed in the related ternary Laves phases with MgZn$_{2}$-type \cite{Laves}. In this section, we, therefore, consider, from ab-initio alloy theory,  the disorder of Co$_3$Mn$_2$Ge, in order to analyze the effect it has on the magnetic state of the material and to compare the results with the theoretical data obtained for the ordered structure. 
The effect of chemical and magnetic disorder on the stability and magnetization of Co$_3$Mn$_2$Ge was investigated as described in the {\it Methods} section (Sec.~\ref{sec:method}). 

 The results of these theoretical calculations is that the ordered structure has a lower total energy, compared to the disordered one. However, configurational entropy contributions to the free energy are reported in the appendix to stabilize the disordered structure at temperatures ~1700 K. The details can be found in {\it Appendix~\ref{sec:acrelax}}.

\begin{figure}[h]
 \centering
 \includegraphics[scale=0.32]{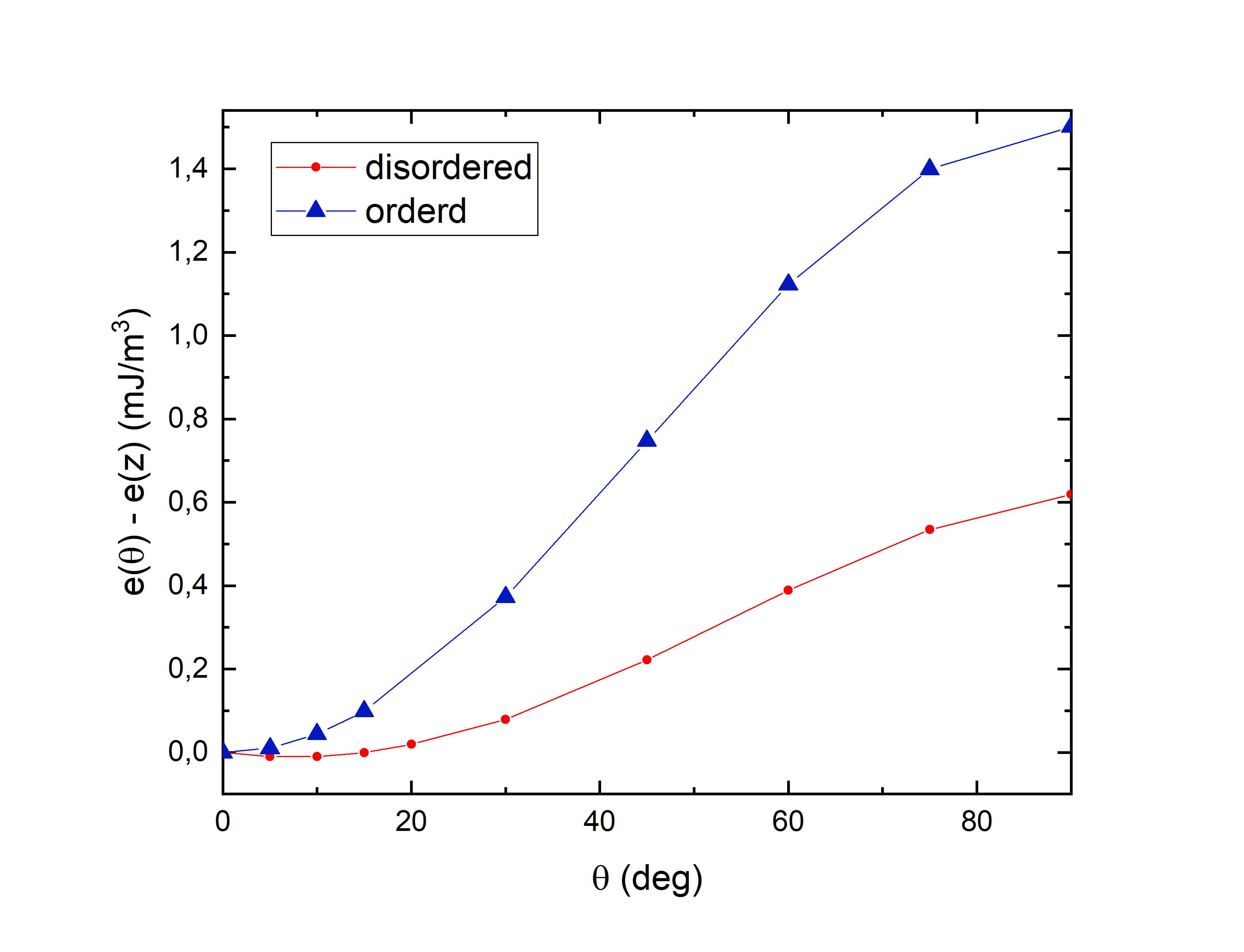}
  \caption{(Color online) MAE for  several magnetization directions ($\phi = 0^{\circ}$) for the ordered Co$_3$Mn$_2$Ge and disordered Mn$_2$(Co$_{0.75}$Ge$_{0.25}$)$_4$ phases.}
 \label{cone}
\end{figure}

Following the experimental findings that point to an easy-cone magnetic anisotropy at the lower temperatures, we calculated MAE for  several magnetization directions ($0^{\circ} < \theta < 90^{\circ}$, $\phi = 0^{\circ}$) for the ordered Co$_3$Mn$_2$Ge and disordered Mn$_2$(Co$_{0.75}$Ge$_{0.25}$)$_4$ phases, see Fig.~\ref{cone}. Indeed, we can see that the disordered system shows a pronounced deviation from the simple uniaxial behavior. Hence, the easy-cone anisotropy observed at lower temperatures is attributed to the Co-Ge disorder observed in the samples. At elevated temperature, the influence of disorder evidently is less pronounced, and the uniaxial anisotropy  that was obtained from theory of ordered samples, is recovered in the experiments.

\section{Chemical substitution of G\lowercase{e}}

In order to find related compounds with similar or improved magnetic properties compared to Co$_3$Mn$_2$Ge, we performed additional theoretical work, where we replaced all Ge atoms in the unit cell by Al, Si, P, Ga, As, In, Sn, Sb, Tl, and Pb. All structures were relaxed and tested for stability by calculating the formation enthalpy with respect to their elemental components (for  details see section {\it Methods}). The systems found to be stable with respect to the elemental components (Tl- and Pb-based materials turned out to be unstable) were investigated further for their magnetic characteristics. Their MAE and the details of the magnetic state are given in Table \ref{table:CoMnGe}.

 \begin{figure}[h]
 \centering
 \includegraphics[scale=0.48]{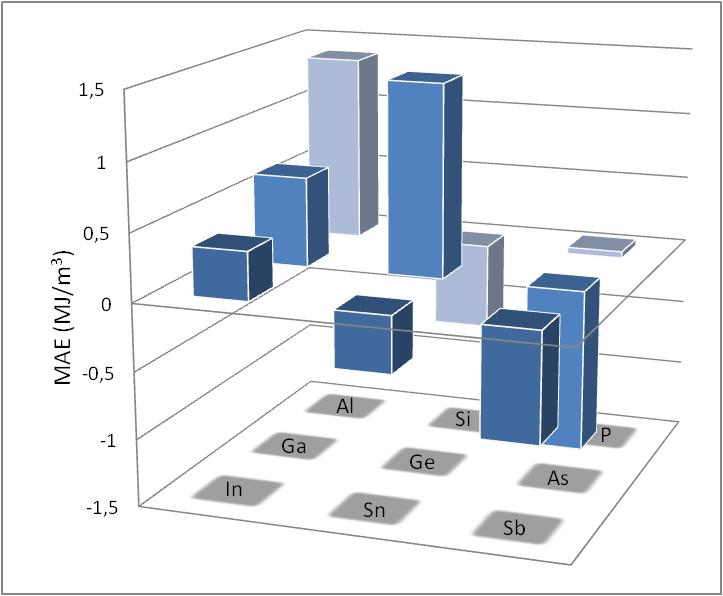}
  \caption{(Color online) Magnetic anisotropy energy of Co$_3$Mn$_2$X, with X = Al, Si, P, Ga, Ge, As, In, Sn, Sb. Positive values indicate a uniaxial anisotrophy.}
 \label{MAE}
\end{figure}

Replacing Ge in Co$_3$Mn$_2$Ge by the neighboring elements, does not change to the magnetic state much, but it has a large effect on the MAE, see Fig.~\ref{MAE}. To determine the origin of the difference in MAE values we analyzed the spin-orbit coupling energy (SOC) of the Co and Mn atoms with their spins oriented along the {\it z} and {\it x} directions as well as the partial density of states (DOS), similar to the analyses presented in Refs.\cite{Arapan,Liu,Antropov,Miura,Mark,Zhang}. The details of that investigation can be found in {\it Appendix}~\ref{sec:CoMnX}. Summarizing the results, we find that apart from the Co$_3$Mn$_2$Ge compound that is listed in the ICSD database,  Co$_3$Mn$_2$Al and Co$_3$Mn$_2$Ga are also expected to have the properties of a good permanent magnet. We have shown that the main contribution to the large MAE of these compounds arises from Co atoms. A synthesis and charaterisation of these two compounds is outside the scope of this investigation, but represents clearly an interesting avenue forward.

\section{Discussion and conclusions}

Using the high-throughput and data-mining approach we filtered through the RE-free materials of the ICSD database that contain necessarily two 3{\it d}-elements.
Only one of those approximately thousand structures satisfied our requirements for a strong permanent magnet,  Co$_3$Mn$_2$Ge, with the saturation magnetization of 1.71 T, MAE equal to 1.44 MJ/m$^3$, and $T_{\rm C}$ of 700 K.
To reduce the cost of this material, we attempted to replace Ge by other elements.
Two materials, Co$_3$Mn$_2$Al and Co$_3$Mn$_2$Ga, were found to have sufficient magnetization, MAE, and Curie temperature to be used as high-performance permanent magnets.
The former of the two can be problematic to produce due to the competing Heusler compound being considerably more stable; the latter doesn't have a similar drawback.

Co$_3$Mn$_2$Ge was successfully synthesized using induction melting followed by annealing at 1073 K. The resulting crystal structure was in good agreement with the previously published results.
However, when the ordering of the system was investigated with precision single crystal analysis, it indicated that the disordered structure type MgZn$_2$ with intermixing Co and Ge on the 6\textit{h} and 2\textit{a} sites was preferred. The structure taken from the ICSD database and used in the high-throughput calculation was, unlike the experimental one, ordered.

Magnetic characteristics of the synthesized compound were measured and produced the saturation magnetization of 0.86 T at the temperature of 10 K, uniaxial magnetic anisotropy equal to 1.18 MJ/m$^3$ above around 175 K (with the easy-cone character of anisotropy below), and $T_{\rm C}$=359 K. These values, even though lower than the theoretical predictions, make Co$_3$Mn$_2$Ge a promising candidate for a high-performance permanent magnet. Further analysis of its magnetic state as well as the ways to tune some of these numbers is highly desirable and represents ongoing work. 

Calculations were also performed for the disordered crystal structure of Co$_3$Mn$_2$Ge obtained in the experiment. We found that the crystallographic ordered structure with FM ordering is more stable than both ferromagnetically and antiferromagnetically ordered configurations of the structurally disordered compound, in a temperature range of 0 -- 1735 K.
Calculations of the magnetocrystalline anisotropy were also performed for the disordered Mn$_2$(Co$_{0.75}$Ge$_{0.25}$)$_4$ phase. These calculations show a pronounced deviation from the uniaxial character of magnetic anisotropy, in agreement with our experimental low temperature data. Combining the results of the theoretical and experimental works, leads to the conclusion that structural disorder influences the magnetic properties, including the MAE, of Co$_3$Mn$_2$Ge in a detrimental way, and that the best properties are expected for the most ordered samples. 

In the {\it Appendix} \ref{sec:comnge_previous} we list the properties of the all the previously known Co--Mn--Ge systems. Based on that information we can conclude, that the disorder is quite common in all the Co--Mn--Ge compounds and depends strongly on the sample preparation and experimental procedures. With that in mind, we believe that further investigation into the sample preparation or some possible doping alternatives is necessary for an attempt to stabilize Co$_3$Mn$_2$Ge in its ordered form, which is expected to have higher magnetization and magnetic anisotropy. Having said that, even the disordered Mn$_2$(Co$_{0.75}$Ge$_{0.25}$)$_4$ phase, with saturation polarization of 0.86 T and uniaxial magnetic anisotropy of 1.18 MJ/m$^3$, possesses the characteristics of a good permanent magnet, although a further investigation into increasing the $T_{\rm C}$ = 359 is desirable.

It is also worth mentioning, that orientation of magnetocrystalline anisotropy and the magnetic moment of Co--Mn--Ge systems is strongly affected by Co:Mn ratio. The synthesized sample of Co$_3$Mn$_2$Ge has the actual composition of Co$_{52}$Mn$_{34}$Ge$_{14}$ which might result in the discrepancy between the magnetic results obtained experimentally and the ones predicted in the high-throughput search. It was proven that the easy-cone magnetocrystalline anisotropy, observed at the temperatures below around 175 K, is the result of the Co--Ge disorder. We would like to point out, however, that one of the best current permanent magnets, Nd$_2$Fe$_{14}$B \cite{Tokuhara1985}, possesses a similar feature. Nevertheless, this fact is another incentive to look for the way of stabilizing the ordered phase of Co$_3$Mn$_2$Ge.



The current investigation is a promising example of the material found in the theoretical data-mining search being synthesized and showing the desired characteristics of a good permanent magnet. Further experimental investigation into the magnetic state of Co$_3$Mn$_2$Ge will be performed as well as the computational search for improving its price-performance. With its Curie temperature close to the room temperature, we can also consider fine-tuning Co$_3$Mn$_2$Ge for magnetocaloric applications.

\section{Methods}\label{sec:method}

\subsection{High-throughput DFT}

The high-throughput screening step to calculate the magnetic moment of the materials was performed using the full-potential linear muffin-tin orbital method (FP-LMTO) including spin-orbit interaction as implemented in the RSPt code \cite{rspt1, rspt2}.
For this step the initial magnetic configuration for all the materials was ferromagnetic (FM).

The magnetic state of materials (FM  or antiferromagnetic - AFM), unless previously known, was determined using Vienna Ab Initio Simulation Package (VASP) \cite{vasp1,vasp2,vasp3,vasp4} within the Projector Augmented Wave (PAW) method \cite{PAW}, along with the Generalized Gradient Approximation (GGA) in Perdew, Burke, and Ernzerhof (PBE) form \cite{PBE}.
VASP was also used for structure relaxation at the post-high-throughput stage as well as to calculate spin-orbit coupling energies, for the analysis presented in the Appendix.

The magnetic anisotropy energy (MAE) was calculated  using the RSPt code as $\Delta E = E^{pl} - E^{c}$; here $E^{c}$ and $E^{pl}$ are the total energies with the magnetization directed along and perpendicular to the $c$-axis.
Calculations were performed with the tetrahedron method with Bl\"ochl correction for the Brillouin zone integration \cite{Blochl}. A positive sign of the MAE corresponds to the required uniaxial anisotropy. For the disordered Co$_3$Mn$_2$Ge, the MAE was evaluated for several polar angles from the force theorem \cite{PhysRevB.41.11919,PhysRevB.32.2115} as the difference of the eigenvalue sums for the two magnetization directions $e^{\theta}$ and $e^{c}$ ($\theta$ is a polar angle while azimuthal angle is equal to zero), while keeping the effective potential fixed. 
From the first principles calculations, the magnetic hardness parameter was evaluated from the expression $\kappa = \sqrt{\Delta{E}/\mu_0 M_S^2}$ \cite{hardness}, where $M_S$ is saturation magnetization and $\mu_0$ is the vacuum permeability.

The Curie temperature $T_{\rm C}$ was calculated using Monte Carlo simulations implemented within the Uppsala  atomistic  spin  dynamics (UppASD) software \cite{ASD}.
Atomistic  spin  dynamics calculations were performed on a $30\times 30\times 30$ supercell with periodic boundary conditions.
The required  exchange parameters were calculated with the RSPt code within the first nine coordination shells \cite{PhysRevB.91.125133}.

Formation enthalpies of the materials were calculated with respect to their elemental components as
\begin{equation*}
\Delta H = H_{\bf Co_3Mn_2X} - 3H_{\bf Co} - 2H_{\bf Mn}-H_{\bf X},
\end{equation*}
for X= Al, Si, P, Ga, As, In, Sn, Sb, Tl, and Pb. In this expression $H_{\bf Co_3Mn_2X}$, $H_{\bf Co}$, $H_{\bf Mn}$, and $H_{\bf X}$ are the enthalpies of formation for Co$_3$Mn$_2$X, hexagonal close-packed (hcp) cobalt, body-centered cubic (bcc) manganese, and element X (such as, for example, cubic close-packed (ccp) aluminum for Co$_3$Mn$_2$Al), respectively.
Formation enthalpies with respect to Heusler alloys Co$_2$MnX were obtained according to the following formula
\begin{equation*}
\Delta H = H_{\bf Co_3Mn_2X} - H_{\bf Co_2MnX} - H_{\bf Co} - H_{\bf Mn},
\end{equation*}
where $H_{\bf Co}$, $H_{\bf Mn}$, and $H_{\bf Co_2MnX}$ are the enthalpies of hcp cobalt, bcc manganese, and the corresponding Heusler alloy.

The effect of chemical and magnetic disorder on the stability and magnetization of Co$_3$Mn$_2$Ge was investigated by means of the coherent potential approximation \cite{Soven1967, Gyorffy1972} as implemented in the Exact Muffin-Tin Orbitals (EMTO) method \cite{Vitos2001, Vitos2007}. We used 
$s$, $p$, $d$ and $f$ orbitals in the basis set.
The one-electron equations were solved within the soft-core and scalar-relativistic approximations. The Green's function was calculated for 16 complex energy points distributed exponentially on a semi-circular contour including states within 1.1 Ry below the Fermi level.
For the one-center expansion of the full charge density a $l_{\rm max}^h$=8 cutoff was used.
The electrostatic correction to the single-site coherent potential approximation was described using the screened impurity model \cite{Korzhavyi1995} with a screening parameter of 0.6.
Total energies were calculated using the PBE \cite{PBE} exchange-correlation functional, while local density approximation \cite{LDA1, LDA2} was used to calculate the magnetic moments and exchange interactions. The latter was calculated within the magnetic force theorem \cite{Liechtenstein1984} for the ferromagnetic and disordered local moment (DLM)\cite{DLM1, DLM2} configurations as implemented in the EMTO code. DLM represents the high temperature paramagnetic (PM) phase. In this model, the paramagnetic phase of Co$_3$Mn$_2$Ge reads as (Co$_{0.5}^\uparrow$Co$_{0.5}^\downarrow$)$_3$(Mn$_{0.5}^\uparrow$Mn$_{0.5}^\downarrow$)$_2$Ge. Similar formulation is applied to the paramagnetic phase of the alloy as well.

\subsection{Synthesis}\label{sec:expmethod}

Samples of Co$_3$Mn$_2$Ge were synthesized by melting Co (Alfa Aesar, 99.9\%), Mn (H\"{o}gan\"{a}s AB, 99.9\%) and Ge (Kurt J. Lesker, 99.999\%) together in an induction furnace under Ar (purity 99.999\%) atmosphere.
The resulting ingots were placed in Al$_2$O$_3$ crucibles, sealed in evacuated quartz glass tubes and annealed at 1073 K for 14 days after which they were quenched in water.
After the final composition had been established by energy dispersive X-ray spectroscopy analyses (see below), starting materials in stoichiometry of Co$_{52}$Mn$_{34}$Ge$_{14}$ were prepared using the established protocol to yield the final samples.
Samples were manually ground and powders taken for analysis. 

\subsection{Crystal structure analysis}

The crystal structure was investigated using X-ray powder diffraction (XRPD), single crystal X-ray diffraction (SCXRD) and scanning electron microscopy (SEM) coupled with energy dispersive X-ray spectrocopy (EDS).
The powders were mounted on single-crystal Si sample holders and X-ray diffraction patterns were collected using a Bruker D8 Advance with monochromatized Cu-K$\alpha{_1}$ ($\lambda$ = 1.540598 \AA) radiation at room temperature.
FullProf was used with the Rietveld refinement method to analyse the data \cite{FullProf}. A Bruker D8 single-crystal X-ray diffractometer with Mo K$\alpha$ radiation ($\lambda$ = 0.71073 \AA) upgraded with an Incoatec Microfocus Source (I$\mu$S, beam size $\approx$ 100$\mu$m at the sample position) and an APEX II CCD area detector (6cm $\times$ 6cm) was utilized to collect SCXRD intensities at room temperature.
SCXRD data reduction and numerical absorption corrections were performed using the APEX III software from Bruker\cite{Apex3}.
The initial model of the crystal structure was first obtained with the program SHELXT-2014 and refined in the program SHELXL-2014 within the APEX III software package.
The microstructure was evaluated with a Zeiss Merlin SEM equipped with a secondary electron (SE) detector and an energy-dispersive X-ray spectrometer.
The samples for electron microscopy analysis were prepared by standard metallographic techniques through grinding with SiC paper. For final polishing a mixture of SiO$_2$ and H$_2$O was used.

\subsection{Magnetic measurements}

Magnetization versus field and temperature measurements were performed using a Quantum Design MPMS XL system. Isothermal magnetization curves were recorded at several temperatures in applied magnetic fields up to 5 T. Magnetization measurements were performed on bulk samples as well as on single crystals.
The single crystals used for these measurement were rather small hexagons, with a height of $100 \mu$m and a side length of $10 \mu$m, resulting in a magnetic moment of $10^{-7}$--$10^{-8}$ Am$^2$. Care was done to avoid common artifacts introduced when using this system \cite{Sawicki2011}. 
The small hexagons were too small to accurately measure the weight of the sample, and thus the hysteresis curves from single crystals were scaled to match the magnetization at the same temperature for bulk samples. 
The temperature dependent magnetization was measured between 10 K and 390 K in the applied magnetic fields of 0.01 T and 1 T.
The temperature dependent magnetization was  also measured between 300 and 900 K and back to 300 K in an LakeShore VSM equipped with a furnace.
The high temperature measurements were performed in an applied magnetic field of 0.01 T using a heating/cooling rate of 3 K/min.
The magnetization in SI units was calculated from the measured magnetic moment by using the sample weight and density obtained from XRD measurements at 298 K.
The law of approach to saturation \cite{Chikazumi,Herbst1998} was used to calculate the effective anisotropy constant of the material, $\left| {K}_{\textrm{eff}} \right|$ assuming the material to be uniaxial.

\section{Acknowledgement}

The authors would like to acknowledge the support of the Swedish Foundation for Strategic Research, the Swedish Energy Agency (SweGRIDS), the Swedish Research Council, The Knut and Alice Wallenberg Foundation, STandUPP and the CSC IT Centre for Science, and the Swedish National Infrastructure for Computing (SNIC) for the computation resources. E. K. D.-Cz. acknowledges A. V. Ruban for valuable discussions. O. Yu. V. acknowledges the support of Sweden's Innovation Agency (Vinnova). 

\section{Contributions}

O.E., H.C.H, P.S. and M.S. initiated the research. A.V. performed the high-throughput search and data analysis. V.S. and S.R.L synthesized the samples. V.S. carried out the crystal structure analysis and SEM/EDS. D.H. performed magnetic measurements. DFT disorder CPA calculations were performed and analysed by E.K.D.-C. and S.H. A.V., V.S., and D.H. drafted the manuscript, which was reviewed and edited by all the authors. All authors contributed to discussions and analysed the data.

\section{Competing interests}

The authors declare no competing interests.

\bibliography{main}

\onecolumngrid
\appendix
\newpage
\section{Materials discarded during the final steps of high-throughput and data-mining search.}

Some of the materials with the magnetic moment above the set-up threshold of 1.0 $\mu_B$/f.u. were discarded later due to the low or planar MAE or saturation magnetization being much lower than 1 T. These materials are listed in Table \ref{table:4}.

\begin{table}[h!]
\centering
\caption{Systems with magnetic moment calculated to be higher than 1.0 $\mu_B$/f.u. during the high-throughput step, which did not fulfill the requirements of a good permanent magnet on the further steps of data-mining. }
\begin{tabular}{l l l l l l } 
 \hline \hline
 Material & ICSD & Space & Mag. & MAE (th) & Sat. magn.  \\ 
  & number & group & state &  MJ/m$^3$ & T \\
 \hline
CrNiAs & 43913 & 189 & FM & 0.37 & 0.86 \\ 
CrNiP & 43913 & 189 & FM & 0.21 & 0.83 \\
Fe$_2$Ti & 103663 & 194 & FM & 0.10 & 0.84 \\
CoCrGe & 409451 & 194 & FM & -0.18 & 0.75 \\
CoGeMn & 623495 & 194 & FM & -0.16 & 1.10 \\
Cu$_2$Co(SnSe$_4$) & 99296 & 121 & FM &  & 0.17 \\
Cu$_2$CoGeS$_4$ & 99293 & 121 & FM &  & 0.24 \\
Cu$_2$CoSiS$_4$ & 99292 & 121 & FM &  & 0.24 \\
Cu$_2$CoSnS$_4$ & 99294 & 121 & FM &  & 0.21 \\
Mn$_2$Co$_2$C & 44353 & 123 & FM & -2.21 & 0.43 \\
MnCoAs & 610084 & 62 & FM & -0.88 & 0.90 \\
MnCoP & 16483 & 62 & FM & -0.41 & 1.00 \\
Ni$_5$Sc & 646468 & 191 & FM &  & 0.20 \\
GeMnSc & 600156 & 189 & FM & 0.24 & 0.51 \\
Sc$_2$FeRh$_5$B$_2$ & 51437 & 127 & FM & -0.13 & 0.39 \\
TiZn$_2$ & 106184 & 194 & FM & -0.01 & 0.30 \\
FeNiAs & 610509 & 189 & FM & -1.53 & 0.64 \\
 \hline \hline
\end{tabular}
\label{table:4}
\end{table} 

\newpage

\section{Crystallographic data}

Here we list the crystallographic and structural data for the hexagonal Co$_{3.24}$Mn$_2$Ge$_{0.76}$ compound along with some of the measurement details.

\begin{table*}[hbt!]
\centering
\caption{Crystallographic data and experimental details of the single crystal structure refinement for the hexagonal Co$_{3.24}$Mn$_2$Ge$_{0.76}$ compound. Measurements were carried out at 296 K with Mo K$\alpha$ radiation.}
\begin{tabular}{l|c} 
 \hline
 \hline
 Empirical formula & Co$_3$Mn$_2$Ge \\ 
 \hline
 Calculated formula & Co$_{3.24}$Mn$_{2}$Ge$_{0.76}$ \\
 Structure type & MgZn$_2$\\
 Formula weight, Mr (g/mol) & 177.92\\
 Space group (No.) & \textit{P}6$_3$/\textit{mmc} (194)\\
 Pearson symbol, \textit{Z} & \textit{hP}12, 4\\
 Unit cell dimensions: & \\
 \textit{a}, \AA & 4.8032(2)\\
 \textit{b}, \AA & 4.8032(2)\\
 \textit{c}, \AA & 7.7378(4)\\
 \textit{V}, \AA & 154.60(1)\\
 Calculated density, $\rho$ (g$\cdot$ cm$^{-3}$) & 7.64\\
 Absorption coefficient, $\mu$ (mm$^{-1}$) & 31.835\\
 Theta range for data collection ($^\circ$) & 4.901 $\div$ 37.119\\
 \textit{F}$(000)$ & 324\\
 Range in \textit{h k l} & -8 $\leq$ {\it h} $\leq$ 7,\\
 & -8 $\leq$ {\it k} $\leq$ 7,\\
 & -13 $\leq$ {\it l} $\leq$ 13,\\
 Total No. of reflections & 3933\\
 No. of independent reflections & 180(\textit{R}$_{eq}$ = 0.0264)\\
 No. of reflections with I $>$ 2$\sigma$(\textit{I}) & 174 (\textit{R}$_{\sigma}$ = 0.0096)\\
 Data/parameters & 180/13\\
 Weighting details & w=1/[$\sigma^2F_o^2$+0.5284P]\\
 & where P=($F_o^2+2F_C^2$)/3\\
 Goodness-of-fit on \textit{F}$^2$ & 1.170\\
 Final \textit{R} indices [\textit{I}$>$2$\sigma$\textit{I}] & \textit{R}$_1$ = 0.0100;\\
 & w\textit{R}$_2$ = 0.0264\\
 \textit{R} indices (all data) & \textit{R}$_1$ = 0.0111;\\
 & w\textit{R}$_2$ = 0.0268\\
 Extinction coefficient & 0.00135(13)\\
 \hline
 \hline
\end{tabular}
\label{table:refine}
\end{table*}

\begin{table*}[h!]
\centering
\caption{Atomic coordinates, anisotropic displacement parameters and selected interatomic distances for the hexagonal Co$_{3.24}$Mn$_{2}$Ge$_{0.76}$ compound.}
\begin{tabular}{l l c c c c} 
 \hline
 \hline
 Atom & Site & \textit{x} & \textit{y} & \textit{z} & \textit{U}$_{eq}$(\AA$^2$) \\
 \hline
 0.837Co1+0.163Ge1 & 6\textit{h} & 0.17153(4) & 0.34306(8) & 1/4 & 0.00594(11)\\
 Mn & 4\textit{f} & 1/3 & 2/3 & 0.56358(7) & 0.00817(16)\\
 0.740Co2+0.260Ge2 & 2\textit{a} & 0 & 0 & 0 & 0.006667(17)\\
 \hline
 & & \textit{U}$_{11}$ & \textit{U}$_{22}$ & \textit{U}$_{33}$ & \textit{U}$_{12}$\\
 \hline
 0.837Co1+0.163Ge1 & & 0.00658(14) & 0.00418(16) & 0.00626(15) & 0.00209(8)\\
 Mn & & 0.00849(18) & 0.00849(18) & 0.0075(2) & 0.00424(9)\\
 0.740Co2+0.260Ge2 & & 0.0079(2) & 0.0079(2) & 0.0041(2) & 0.00397(10)\\
 \hline
 Atoms 1,2 & \multicolumn{2}{c}{{\it d} 1,2(\AA)} & Atoms 1,2 & \multicolumn{2}{c}{{\it d} 1,2 (\AA)} \\
 \hline
 2Co$\vert$Ge1--Co2$\vert$Ge2 & \multicolumn{2}{c}{2.4039(2)} & 3Co1$\vert$Ge1--Mn1 & \multicolumn{2}{c}{2.7978(3)} \\
 2Co1$\vert$Ge1--Mn1 & \multicolumn{2}{c}{2.7815(5)} & 5Co2$\vert$Ge2--Mn1 & \multicolumn{2}{c}{2.8178(2)}\\
 \hline
 \hline
\end{tabular}
\label{table:7}
\end{table*}

\textit{U}$_{eq}$ is defined as one third of the trace of the orthogonalized \textit{U}$_{ij}$ tensor. \textit{U}$_{13}$,\textit{U}$_{23}$ = 0.

\section{Ordering temperature, magnetic configuration, and crystal structure of the disordered C\lowercase{o}$_3$M\lowercase{n}$_2$G\lowercase{e}}\label{sec:acrelax}

\begin{figure*}[hbt!]
     \centering
     \begin{subfigure}[b]{0.46\textwidth}
         \centering
         \includegraphics[width=\textwidth]{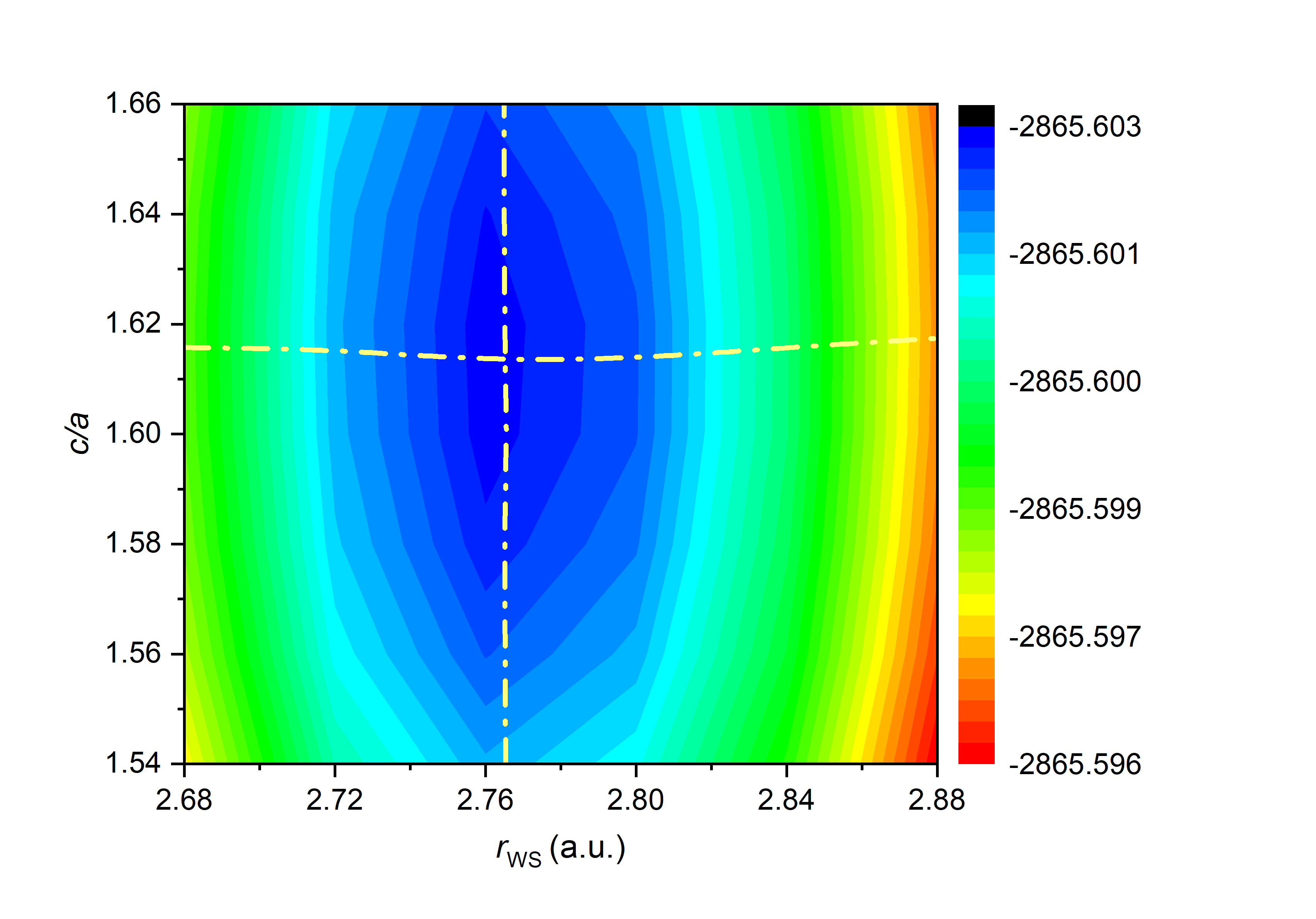}
         \label{Order_S}
     \end{subfigure}
     \hfill
     \begin{subfigure}[b]{0.46\textwidth}
         \centering
         \includegraphics[width=\textwidth]{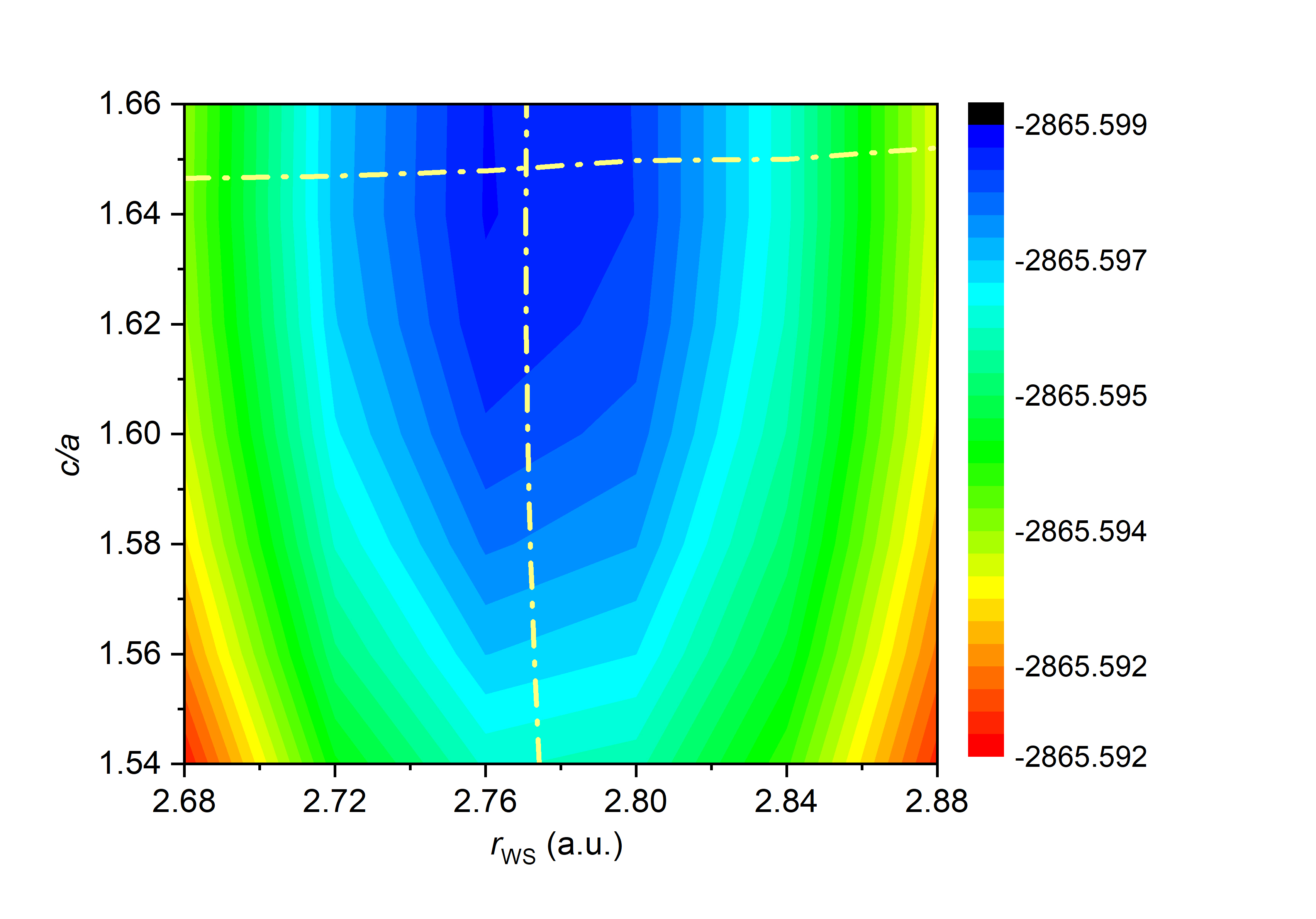}
         \label{Disorder_S}
     \end{subfigure}
        \caption{(Color online) Volume and $c/a$ relaxation of the ordered Co$_3$Mn$_2$Ge (left) and disordered (right) Mn$_2$(Co$_{0.75}$Ge$_{0.25}$)$_4$ structures in the FM phase. $r_{\rm WS}$ denotes the Wigner-Seitz radius. Yellow lines follow the minimum of E($r_{\rm WS}$) curve for a specific value of $c/a$ and the minimum of $E(c/a)$ curve for each $r_{\rm WS}$.}
        \label{acColor}
\end{figure*}

\begin{figure*}[hbt!]
     \centering
     \begin{subfigure}[b]{0.46\textwidth}
         \centering
         \includegraphics[width=\textwidth]{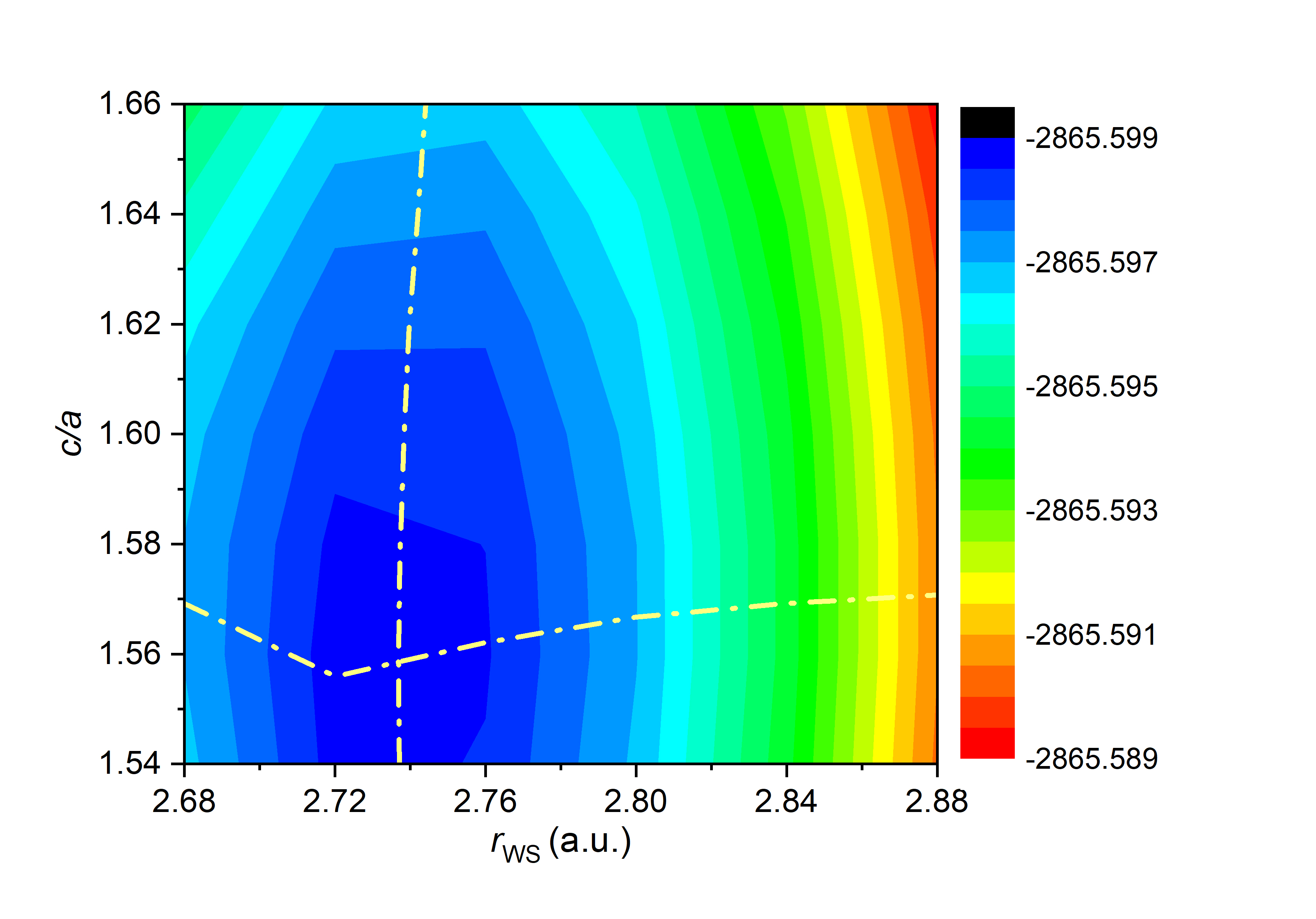}
         \label{Order_S2}
     \end{subfigure}
     \hfill
     \begin{subfigure}[b]{0.46\textwidth}
         \centering
         \includegraphics[width=\textwidth]{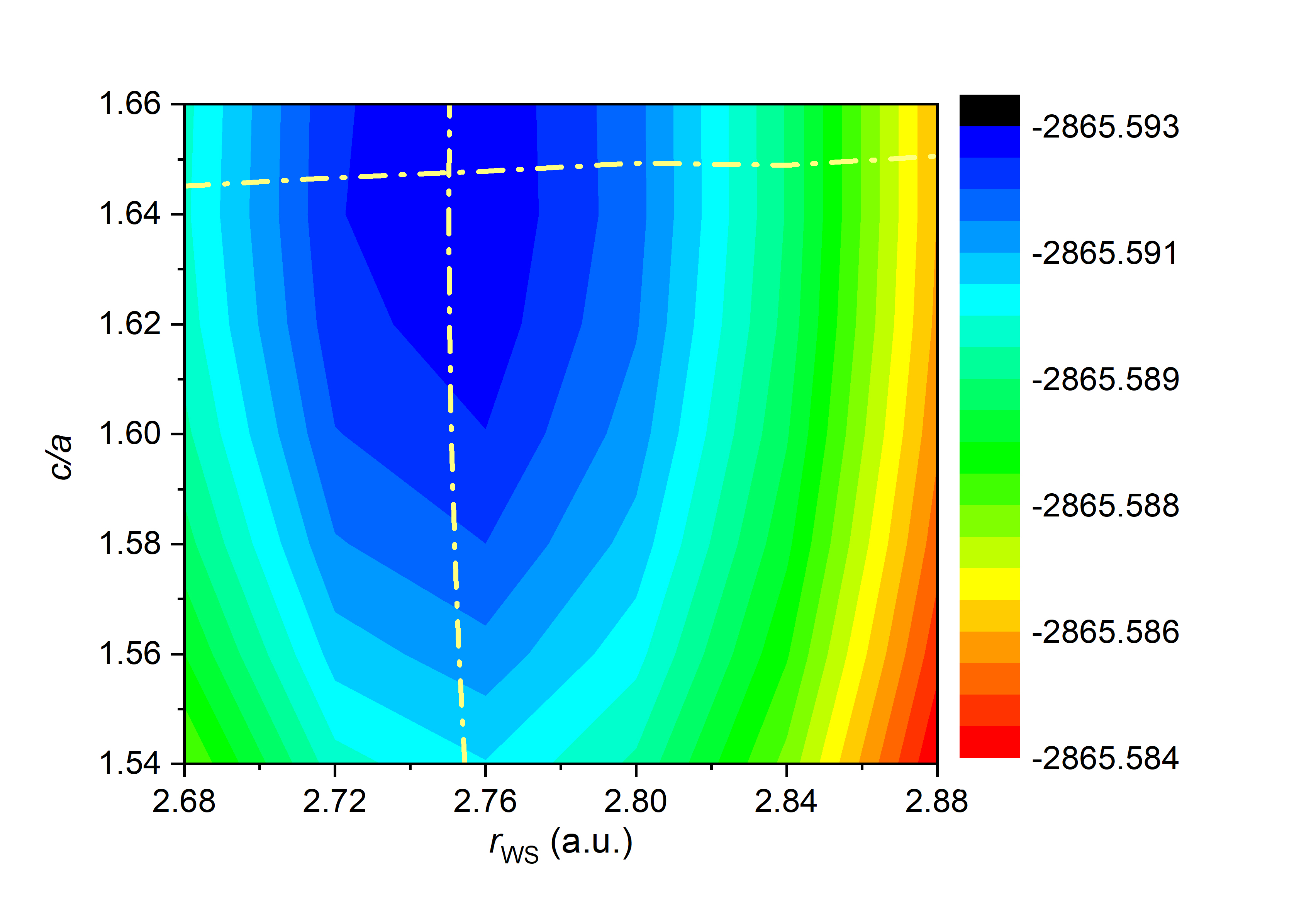}
         \label{Disorder_S2}
     \end{subfigure}
        \caption{(Color online) Volume and $c/a$ relaxation of the ordered Co$_3$Mn$_2$Ge (left) and disordered (right) Mn$_2$(Co$_{0.75}$Ge$_{0.25}$)$_4$ structures in PM phase. $r_{\rm WS}$ denotes the Wigner-Seitz radius. Yellow lines follow the minimum of E($r_{\rm WS}$) curve for a specific value of $c/a$ and the minimum of $E(c/a)$ curve for each $r_{\rm WS}$.}
        \label{acColor2}
\end{figure*}

In this section, we present the additional data for the ordered Co$_3$Mn$_2$Ge and disordered Mn$_2$(Co$_{0.75}$Ge$_{0.25}$)$_4$ structures relaxed with EMTO code in the FM (Fig.~\ref{acColor}) and PM (Fig.~\ref{acColor2}) phases. Table~\ref{table:Disor} lists the crystal structure and magnetic parameters obtained for these structures for both FM and PM magnetic configurations. Fig.~\ref{disorder} summarizes the temperature effects for the FM and PM phases for the c/a-ratio fixed to the experimental value. 

To calculate the temperature which makes the ordered and disordered state degenerate in the FM and PM phase, the crystal structures of Co$_3$Mn$_2$Ge and Mn$_2$(Co$_{0.75}$Ge$_{0.25}$)$_4$ were relaxed in volume and c/a ratio using the EMTO code, as shown in Fig.~\ref{acColor} and Fig.~\ref{acColor2}. Table~\ref{table:Disor}  lists the crystal structure and magnetic parameters obtained  for these states for both FM and PM magnetic configurations. As can be seen, the chemical disorder does not have significant effect on the magnetic properties in the FM state. However, magnetic configuration (FM or DLM) does strongly affect the atomic magnetic moments, especially in the case of Co (Table~\ref{table:Disor}).

The order-disorder transition can be estimated by the cross point of the free energies of the ordered and disordered structures, i.e. $\Delta F$ = $F_{\rm dis}$-$F_{\rm ord}$. The temperature dependent free energy of the FM phase $F(T)^{\rm FM}$ is estimated as $F(T)^{\rm FM}\approx E_0^{\rm FM}+F_{\rm conf}$, where $E_0^{\rm FM}$ is the internal energy for FM state at 0 K and $F_{\rm conf}$ is configurational free energy evaluated at different temperatures. The energy difference of Co$_3$Mn$_2$Ge between the FM ordered and FM disordered (Mn$_2$(Co$_{0.75}$Ge$_{0.25}$)$_4$) states is about 4.124 mRy; the difference in the configurational entropy of the alloy between the ordered and disordered phases equals to 0.375 $k_{\rm B}$. Since $F_{\rm conf}$ = -$TS_{\rm conf}$ we find the transition temperature to be around 1735 K in the FM state.


Temperature effects are summarized for the FM and PM phases in Figure~\ref{disorder} for $c/a$-ratio fixed to the experimental value. As we can see, there is no transition up to either the measured (359 K) or the theoretically predicted (700 K) magnetic transition temperature (see left panel of Fig.~\ref{disorder}). The ordered and disordered structures become degenerate in energy at $\approx$ 2300 K in the PM phase (see right panel of Fig.~\ref{disorder}). The transition temperature is around 2200 K in the PM state. The stability of various antiferromagnetic configurations was investigated as well, for both ordered and disordered case (not shown), to account for the discrepancy between experimental and theoretical $T_{\rm C}$ for fixed $c/a$-ratio. None of the antiferromagnetic states considered in the calculations are stabilized by the configurational entropy up to the $T_{\rm C}$.

The Curie  temperature  for  the  ordered  system  and  for Mn$_2$(Co$_{0.75}$Ge$_{0.25}$)$_4$ was estimated from $J_{ij}$’s calculated for the ferromagnetic reference state and performing subsequent Monte-Carlo (MC) simulations. We obtained 720~K for Co$_3$Mn$_2$Ge and 760 K for the disordered case using the theoretical lattice parameters given in Table \ref{table:Disor}. It is satisfying to note that the $T_{\rm C}$ for the
ordered system, obtained with EMTO method, is in good agreement with the RSPt results.  The Curie temperature
increase with disorder is due to the decrease of the antiferromagnetic Mn-Mn nearest neighbor interactions as chemical disorder is applied. $T_{\rm C}$ estimated for the experimental structure and composition given in Table~\ref{table:SCXRD Ref} is 820 K. We can conclude that the
discrepancy between the predicted $T_{\rm C}$ and the measured
one  does  not  come  from  the  order-disorder  effect alone. To address  this  issue, DLM  calculations  were also performed. These result in the considerable drop of Co local  magnetic  moment  in the DLM state compared to the FM moments, while Mn moment does not change (see Table~\ref{table:Disor}). Magnetic moment of any atomic species, that is reduced in the DLM configuration compared to a FM configuration, will lead to the reduced strength of the inter-atomic exchange interactions, and  reduce of the ordering temperature.

\begin{table*}[h!]
\centering
\caption{Optimized crystal structure parameters ($a$, $c/a$, and volume) and magnetic moments per unit cell calculated for the ordered Co$_3$Mn$_2$Ge and disordered Mn$_2$(Co$_{0.75}$Ge$_{0.25}$)$_4$ phases using the EMTO code, along with the low temperature experimental lattice parameters and magnetic saturation field. Third column contains experimental data. Fourth column (labeled ''Co-excess'') contains experimental data for the low temperature structure, and calculated magnetic moments for this structure and for a composition given in Table~\ref{table:SCXRD Ref} (54 \% Co). }
\begin{tabular}{l l l l l } 
 \hline \hline
  & Ordered & Disordered & Experiment&Co-excess  \\ 
\hline
FM&&&&\\
$a$, $\mathrm{\mathring{A}}$ &  4.83 & 4.81 & 4.803&4.803 \\
$c/a$ &  1.61 & 1.65 & 1.611&1.611 \\
$V$, $\mathrm{\mathring{A}}^3$ & 157.6 & 158.5 & 154.6&154.5 \\
$m_{\rm Co}^{6h}$, $\mu_{\rm B}$ & 1.60 & 1.59 &&1.61 \\
$m_{\rm Mn}$, $\mu_{\rm B}$ & 3.30 & 3.28 &&3.27 \\
$m_{\rm Co}^{2a}$, $\mu_{\rm B}$ &  & 1.56 &&1.63 \\
$M_{\rm tot}$/cell, T&1.66  &1.65 &0.86&1.75\\
\hline
DLM&&&\\
$a$, $\mathrm{\mathring{A}}$ &  4.84 & 4.77 && 4.803\\
$c/a$ &  1.56 & 1.65 & &1.611\\
$V$, $\mathrm{\mathring{A}}^3$ & 152.7 & 154.9 &&154.5 \\
$m_{\rm Co}^{6h}$, $\mu_{\rm B}$ & 0.04 & 0.41 &&0.60 \\
$m_{\rm Mn}$, $\mu_{\rm B}$ & 2.92 & 2.99 &&3.05 \\
$m_{\rm Co}^{2a}$, $\mu_{\rm B}$ &  & 0.05 && 0.51\\
 \hline \hline
\end{tabular}
\label{table:Disor}
\end{table*}

\begin{figure*}[hbt!]
     \centering
     \begin{subfigure}[b]{0.5\textwidth}
         \centering
         \includegraphics[width=\textwidth]{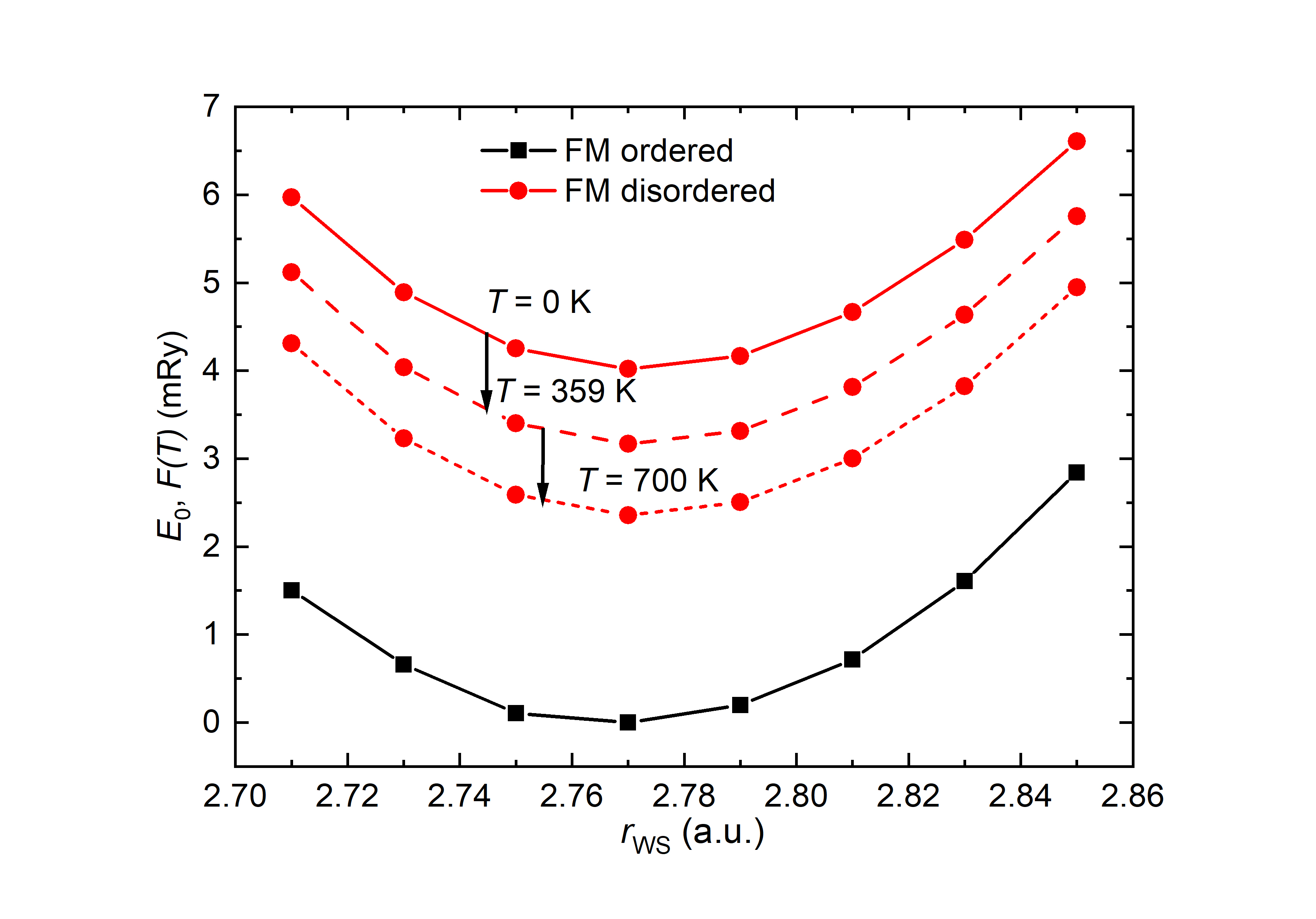}
         \label{energy0k}
     \end{subfigure}
     \hfill
     \begin{subfigure}[b]{0.46\textwidth}
         \centering
         \includegraphics[width=\textwidth]{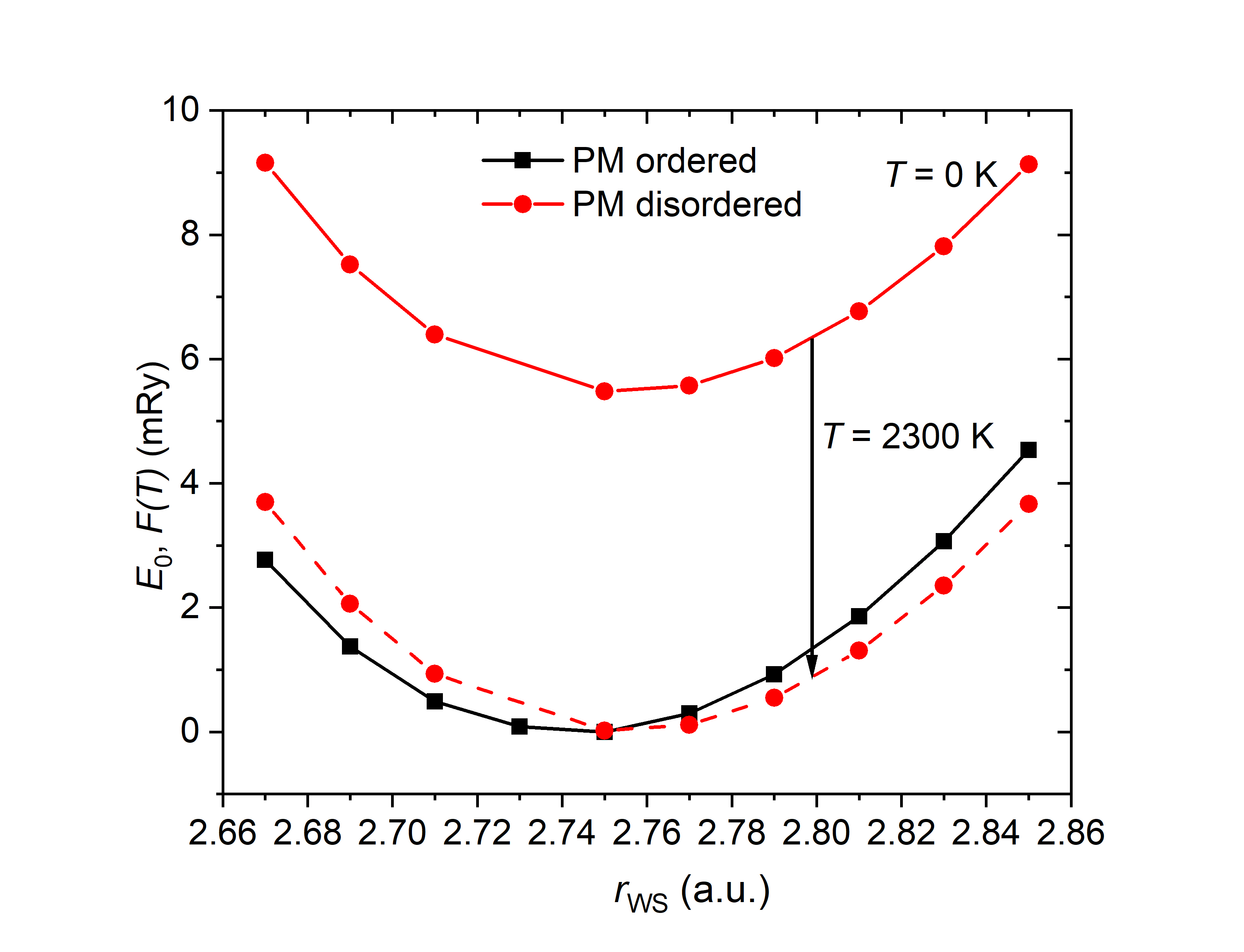}
         \label{energy1750k}
     \end{subfigure}
        \caption{(Color online) Left panel: Normalized total energies ($E_0$) (full lines) and free energies ($F (T)$) (dashed and dotted lines) calculated for the ordered (black squares) and disordered (red circles) FM state of Co$_3$Mn$_2$Ge at 0 K, 359 K (experimental $T_{\rm C}$) and  700 K (theoretical $T_{\rm C}$). Right panel: Normalized total energies ($E_0$) (full lines) and free energies ($F (T)$) calculated for the ordered (black squares) and disordered (red circles)  Co$_3$Mn$_2$Ge at 0 K and 2300 K in the PM state. Dashed line stands for $F(T)$= $E_0$ + $F_{\rm conf}$, dotted line denotes $F(T)$= $E_0$ + $F_{\rm conf}$ + $F_{\rm mag}$. On the x-axis, $r_{\rm WS}$ denotes the Wigner-Seitz radius calculated as $r_{\rm WS}$ = $\sqrt[3]{3V_{\rm atom}/4/\pi}$, where $V_{\rm atom}$ is the average volume of an atom in the unit cell. $F_{\rm conf}$= -$k_{\rm B}T\sum_i$c$_i$ln(c$_i$) and $F_{\rm mag}$=-$k_{\rm B}T$$\sum_i$c$_i$ln(1+$m_i$). }
        \label{disorder}
\end{figure*}


\section{Origin of magnetic anisotropy in C\lowercase{o}$_3$M\lowercase{n}$_2$G\lowercase{e}}\label{sec:CoMnX}

The magnetocrystalline anisotropy originates from the SOC, since it is the only term in the Hamiltonian that couples spin- and real-space,  
something that was first suggested by Van Vleck \cite{VanVleck}. In the case of transition metals where SOC is much smaller than the bandwidth or crystal field, it can be treated as a perturbation. This lead to the possibility to connect the MAE with anisotropy in orbital moment \cite{Bruno}. The original expression for this coupling, made by Bruno, was subsequently extended, approximated, and used for various applications in \cite{Laan, Antropov,Kosugi,Zhang,Mark,Andersson,Alex,Wang}. 

 \begin{figure}[h]
 \centering
 \includegraphics[scale=0.32]{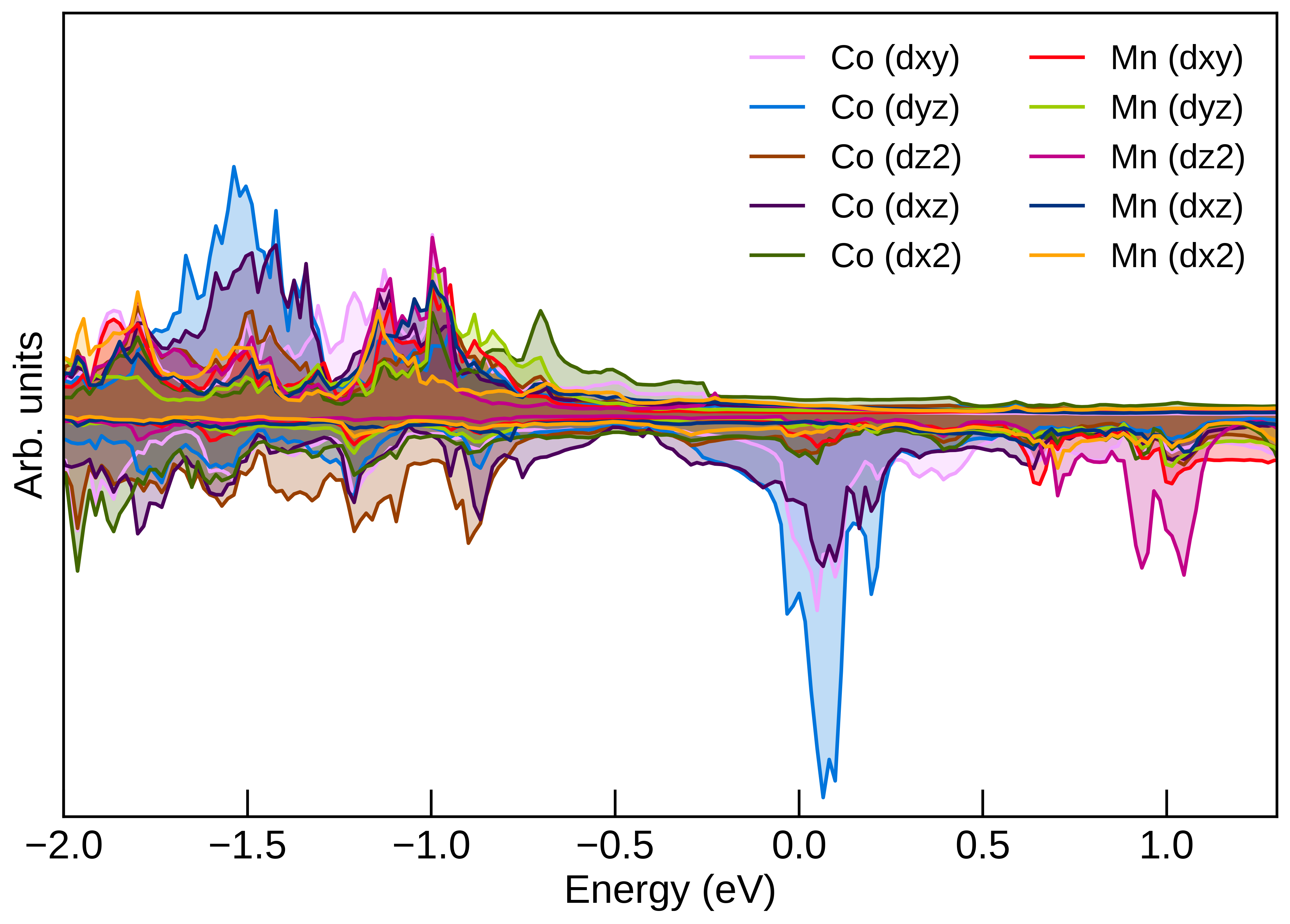}
  \caption{(Color online) Contribution to the  density of states of Co$_3$Mn$_2$Ge from the $d_{x^2-y^2}$ and $d_{z^2}$ ($e_g$ set); $d_{xy}$, $d_{yz}$, and $d_{xz}$ ($t_{2g}$ set) orbitals without SOC interaction. Fermi energy is set at zero.}
 \label{DOS_Ge}
\end{figure}

 \begin{figure}[h]
 \centering
 \includegraphics[scale=0.32]{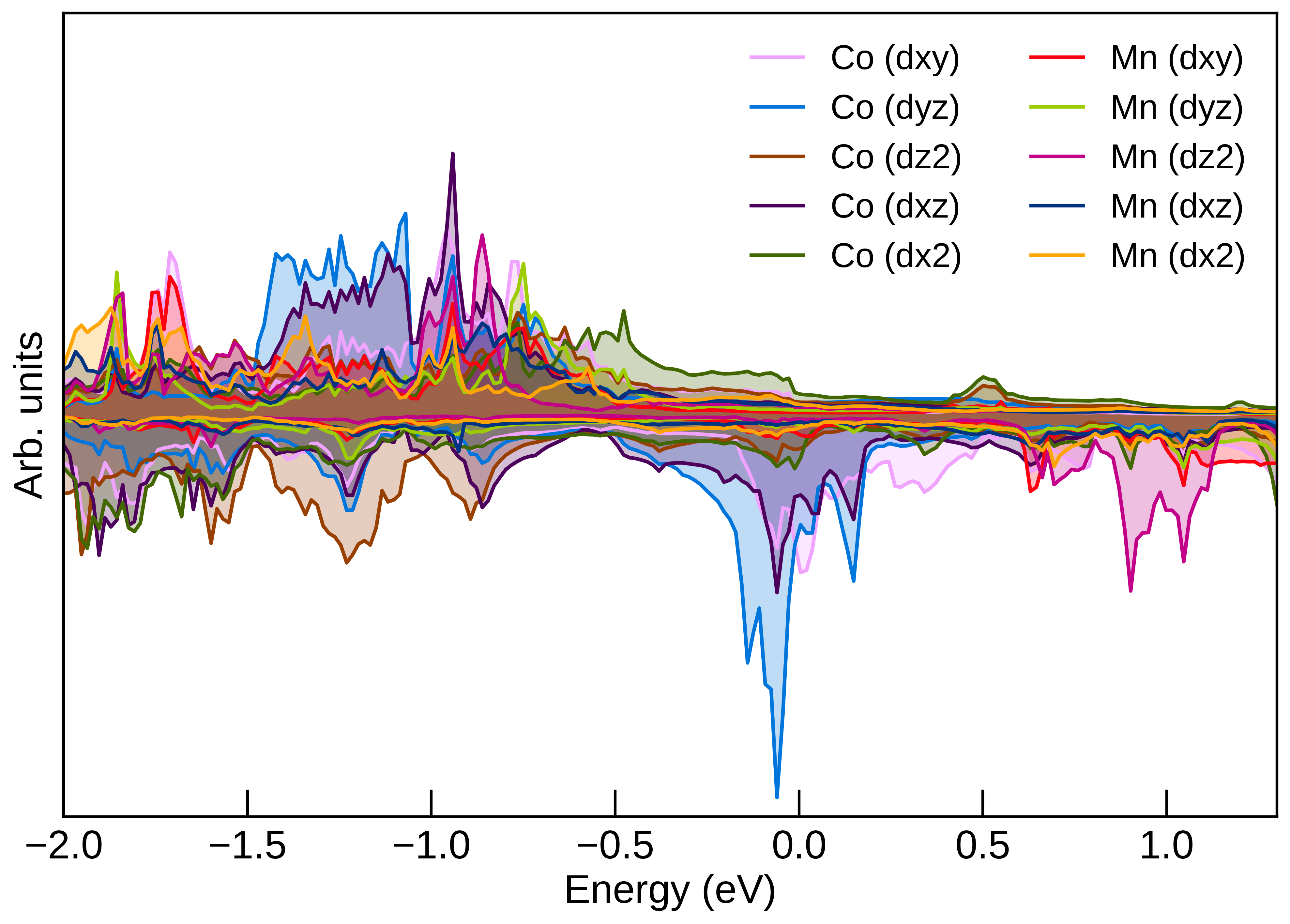}
 \caption{(Color online) Contribution to the  density of states of Co$_3$Mn$_2$As from the $d_{x^2-y^2}$ and $d_{z^2}$ ($e_g$ set); $d_{xy}$, $d_{yz}$, and $d_{xz}$ ($t_{2g}$ set) orbitals without SOC interaction.}
 \label{DOS_As}
\end{figure}

 \begin{table*}[t]
\centering
\caption{Calculated MAE (RSPt), saturation magnetization (RSPt), the average magnetic moment per Co and Mn atoms, Curie temperature (UppASD), formation enthalpies with respect to the elemental components for the relaxed Co$_3$Mn$_2$X structures ($\Delta H_{el}$), and formation enthalpies with respect to the Co$_2$MnX Heusler compounds (X = Al, Si, P, Ga, Ge, As, In, Sn, Sb, Tl, and Pb) (VASP), marked $\Delta H_{Heus}$. The last three columns contain the difference in SOC energy with spin orientation along {\it z} and  {\it x} axis (negative sign corresponds to uniaxial magnetic anisotropy) for Mn and Co atoms (VASP). Subscript $_2$ notes two cobalt atoms positioned at (0.667, 0.833, 0.75) and (0.333, 0.166, 0.25) 6{\it h} sites, subscript $_1$ points at the remaining four Co atoms.}
\begin{tabular}{l l l l l l l l c c c} 
 \hline \hline
 Material & MAE & Sat. & M (Co) & M (Mn) & $T_{\bf C}$ & $\Delta H_{el}$ & $\Delta H_{Heus}$ & $\Delta E_{so}$(Mn) & $\Delta E_{so}$(Co$_{1}$) & $\Delta E_{so}$(Co$_{2}$) \\ 
   &  MJ/m$^3$ & magn., T & $\mu_B$ & $\mu_B$ & K & eV/f.u. & eV/f.u. & meV & meV & meV \\
 \hline
Co$_3$Mn$_2$Al & 1.38 & 1.77 & 1.45 & 3.37 & 820 & -1.78 & 0.274 & 0.016 & -0.42 & -0.46 \\ 
Co$_3$Mn$_2$Si & -0.64 & 1.63 & 1.14 & 3.32 & & -2.34 & 0.090 & -0.057 & 0.11 & 1.20 \\
Co$_3$Mn$_2$P & 0.047 & 1.50 & 0.76 & 3.35 & & -2.62 & & 0.018 & -0.04 & 0.10  \\
Co$_3$Mn$_2$Ga & 0.67 & 1.75 & 1.50 & 3.47 & 800 & -1.49 & 0.061 & 0.100 & -0.25 & -0.30 \\
Co$_3$Mn$_2$Ge & 1.44 & 1.71 & 1.44 & 3.52 & 700 & -1.57 & 0.070 & 0.005 & -0.26 & 0.13 \\
Co$_3$Mn$_2$As & -1.2 & 1.58 & 1.08 & 3.50 & & -1.60 & & -0.011 & 0.01 & 1.13 \\
Co$_3$Mn$_2$In & 0.36 & 1.65 & 1.60 & 3.62 & & -0.14 & & 0.160 & -0.11 & -0.25 \\
Co$_3$Mn$_2$Sn & -0.42 & 1.63 & 1.53 & 3.59 & & -0.57 & & 0.064 & -0.21 & 0.03 \\
Co$_3$Mn$_2$Sb & -0.81 & 1.56 & 1.35 & 3.56 & & -0.67 & & 0.047 & -0.19 & 0.68  \\
Co$_3$Mn$_2$Tl & -0.21 & 1.63 & 1.63 & 3.67 & & 0.87 & & & & \\
Co$_3$Mn$_2$Pb & -2.7 & 1.58 & 1.58 & 3.66 & & 0.93 & & & & \\
 \hline \hline
\end{tabular}
\label{table:CoMnGe}
\end{table*}

It has been shown previously \cite{Alex,Laan,Mark,Zhang} that the coupling between the occupied and unoccupied states close to the Fermi energy $\epsilon_F$ dominates the spin-orbit induced change of the total energy. 
Matrix elements $\bra{\mu\sigma}L\cdot S \ket{\mu'\sigma'}$ \cite{Abate,Takayama} determine the spin quantization axis direction which modifies the eigenvalues of the Kohn-Sham Hamiltonian. In this expression, $\mu$ represents a d-orbital (with symmetry $\{ xy, yz, zx, x^2-y^2,z^2r\}$), $\sigma$ denotes spin, while $L$ and $S$ are orbital and spin angular momentum operators. For states within the same spin channel, the couplings $d_{xy} \xrightarrow{}d_{x^2-y^2}$, $d_{yz} \xrightarrow{}d_{xz}$  promote the uniaxial anisotropy, while $d_{xy} \xrightarrow{} d_{xz}$, $d_{xy} \xrightarrow{} d_{yz}$,  $d_{xz} \xrightarrow{} d_{z^2}$, $d_{yz} \xrightarrow{} d_{z^2}$, $d_{xz} \xrightarrow{}d_{x^2-y^2}$, and $d_{yz} \xrightarrow{}d_{x^2-y^2}$ favour the easy-plane magnetocrystalline anisotropy \cite{Alex,Mark,Zhang,Kosugi,Antropov,Laan}. The situation is reversed for the couplings between opposite spin channels. Table \ref{table:5} (in {\it Appendix~\ref{sec:xyz}}) lists the transitions that contribute either to easy-plane or uniaxial anisotropy along with their relative weights. 

\begin{figure}[hbt!] 
\begin{subfigure}{0.4\textwidth}
\includegraphics[width=\linewidth]{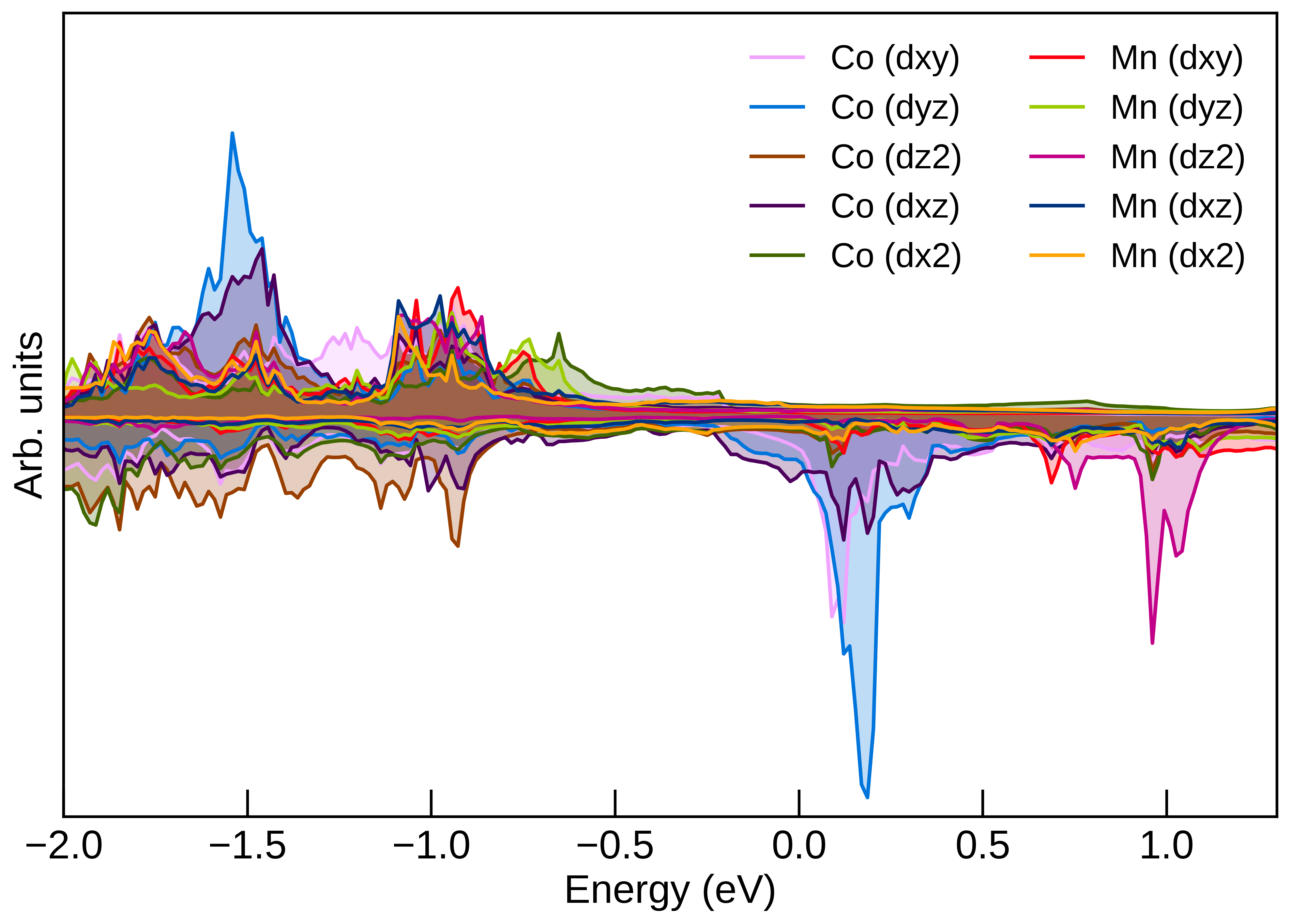}
\caption{Density of states of Co$_3$Mn$_2$Al (MAE = 1.38 MJ/m$^3$).} \label{DOS-Al}
\end{subfigure}\hspace*{\fill}
\begin{subfigure}{0.4\textwidth}
\includegraphics[width=\linewidth]{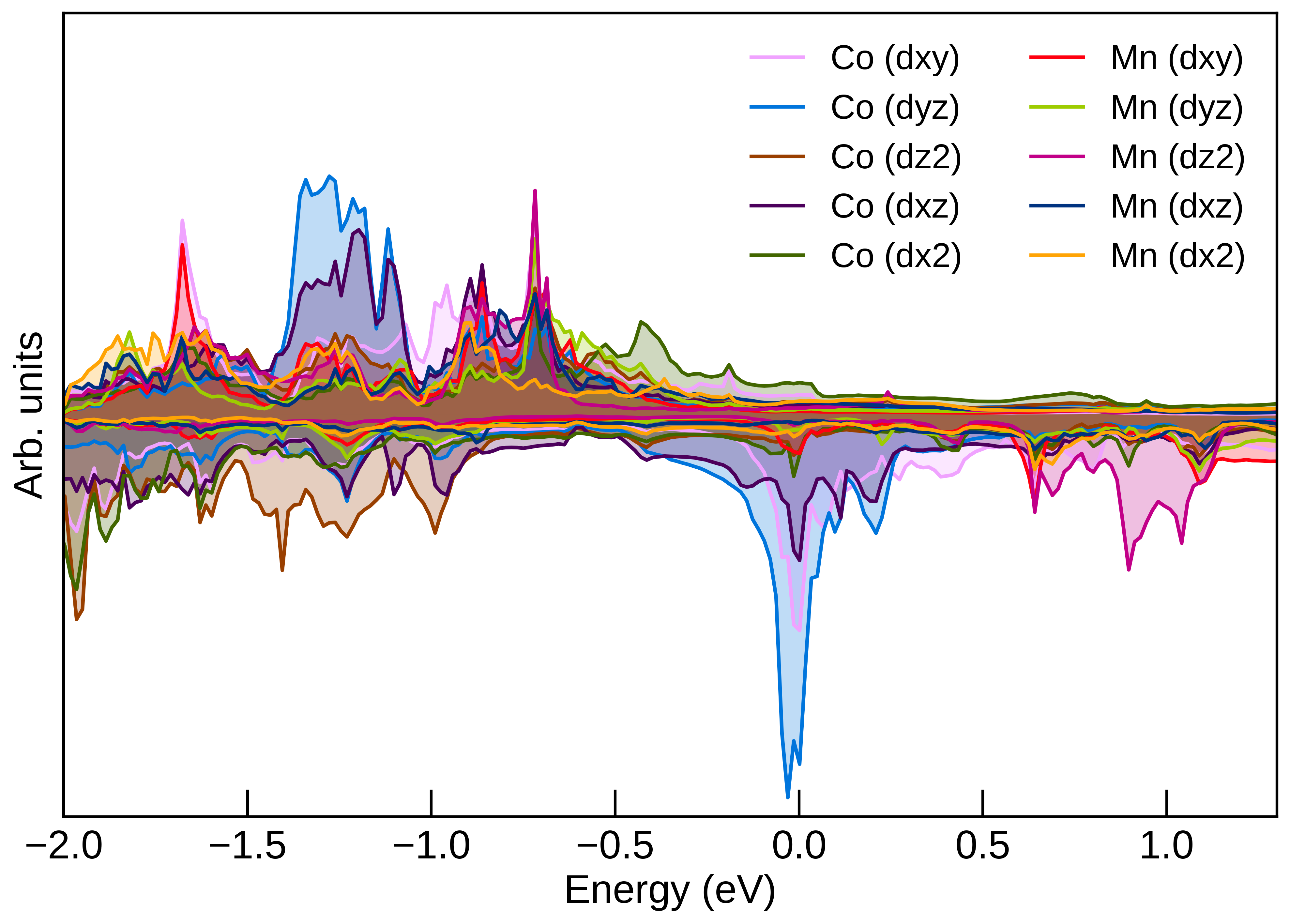}
\caption{Density of states of Co$_3$Mn$_2$Si (MAE = -0.64 MJ/m$^3$).} \label{DOS-Si}
\end{subfigure}

\medskip
\begin{subfigure}{0.4\textwidth}
\includegraphics[width=\linewidth]{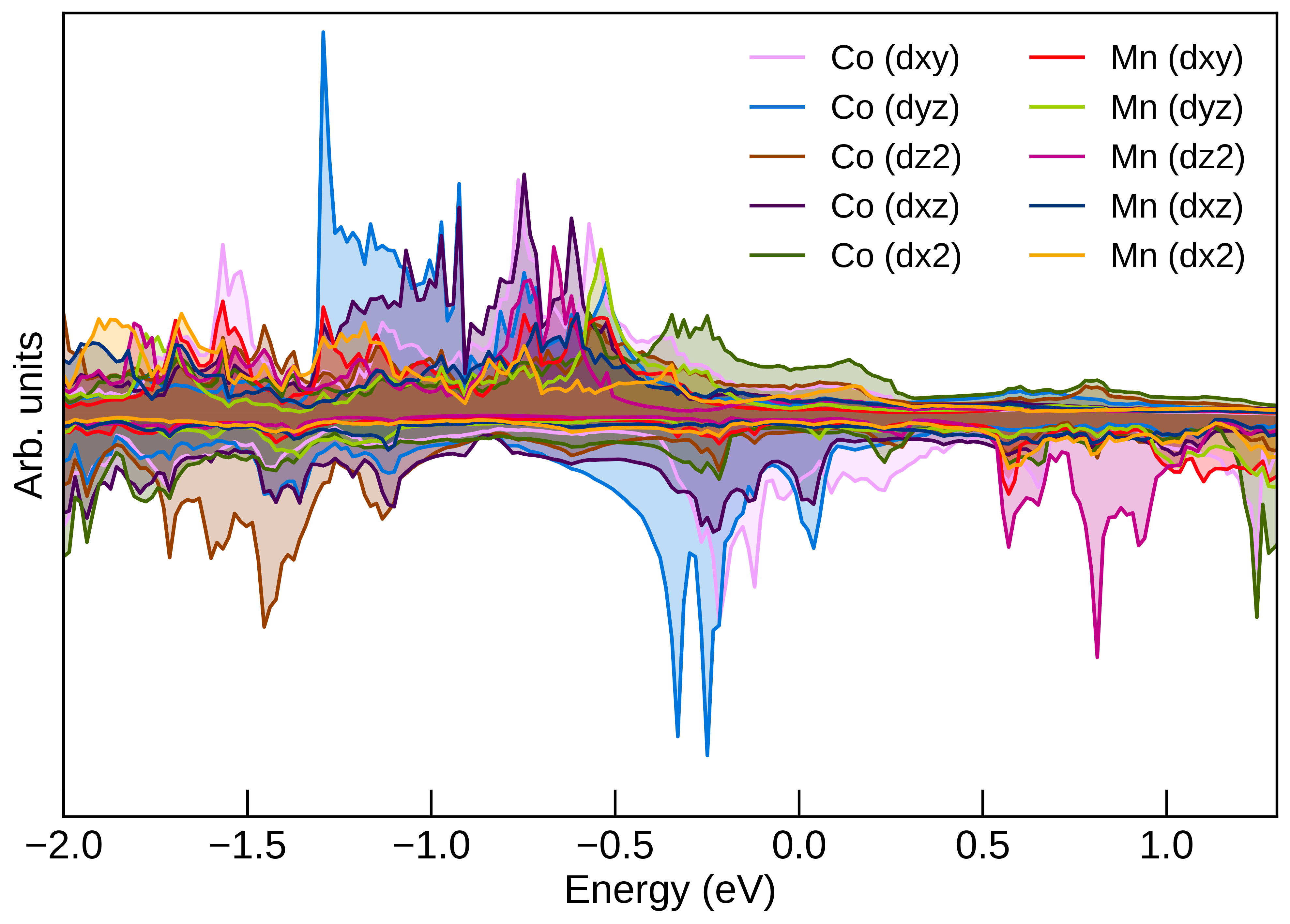}
\caption{Density of states of Co$_3$Mn$_2$P (MAE = 0.047 MJ/m$^3$).} \label{DOS-P}
\end{subfigure}\hspace*{\fill}
\begin{subfigure}{0.4\textwidth}
\includegraphics[width=\linewidth]{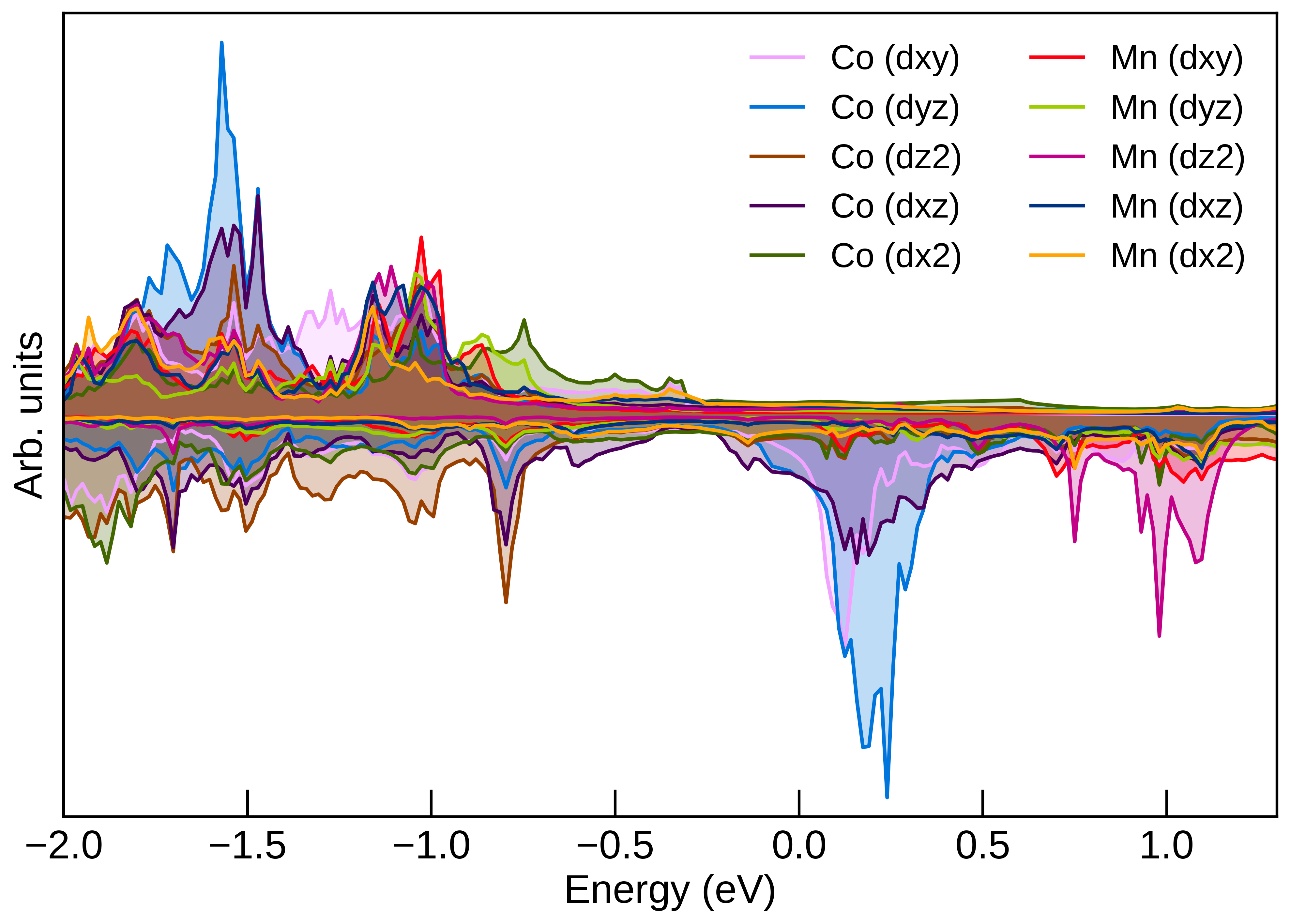}
\caption{Density of states of Co$_3$Mn$_2$Ga (MAE = 0.67 MJ/m$^3$).} \label{DOS-Ga}
\end{subfigure}

\medskip
\begin{subfigure}{0.4\textwidth}
\includegraphics[width=\linewidth]{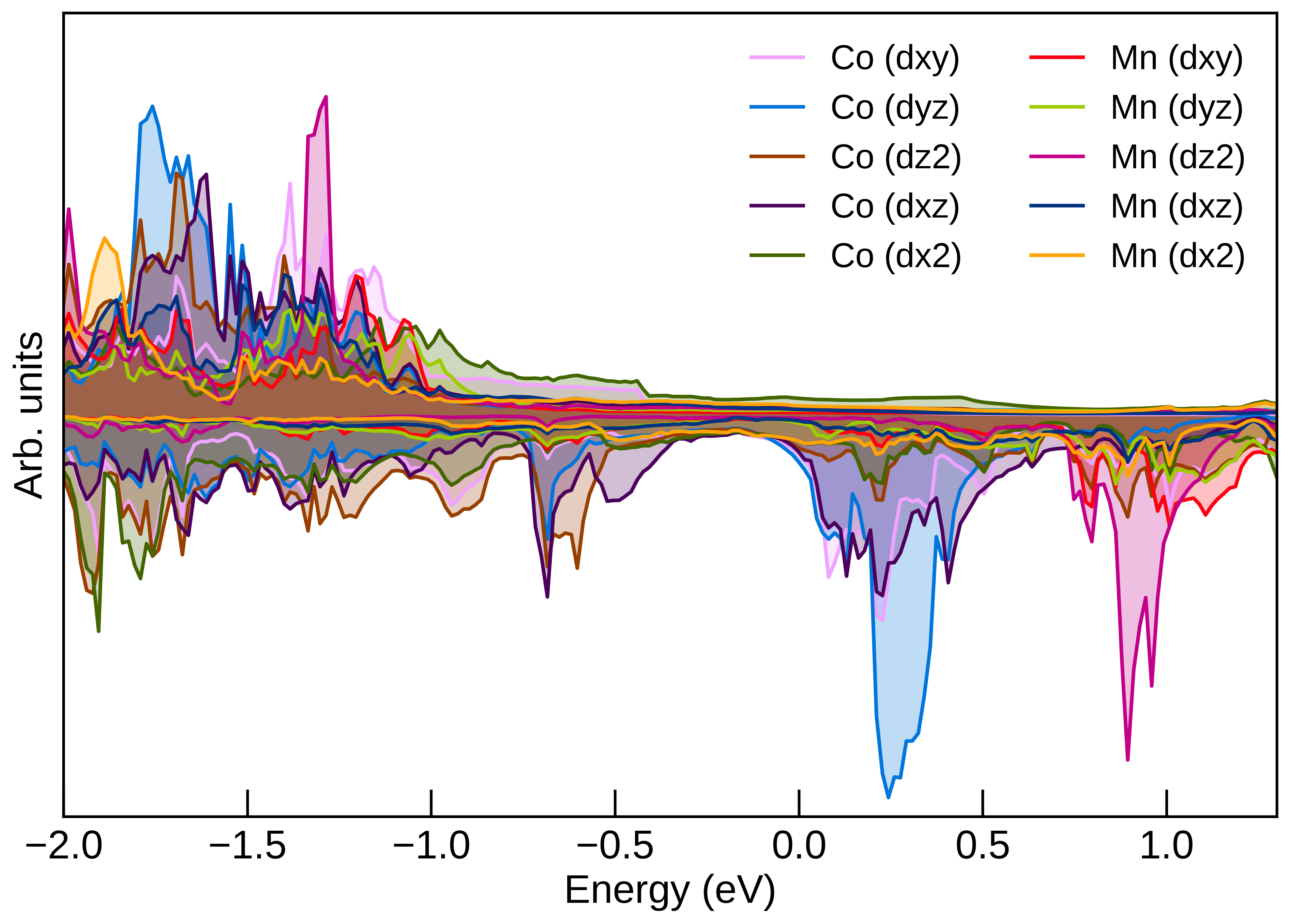}
\caption{Density of states of Co$_3$Mn$_2$In (MAE = 0.36 MJ/m$^3$).} \label{DOS-In}
\end{subfigure}\hspace*{\fill}
\begin{subfigure}{0.4\textwidth}
\includegraphics[width=\linewidth]{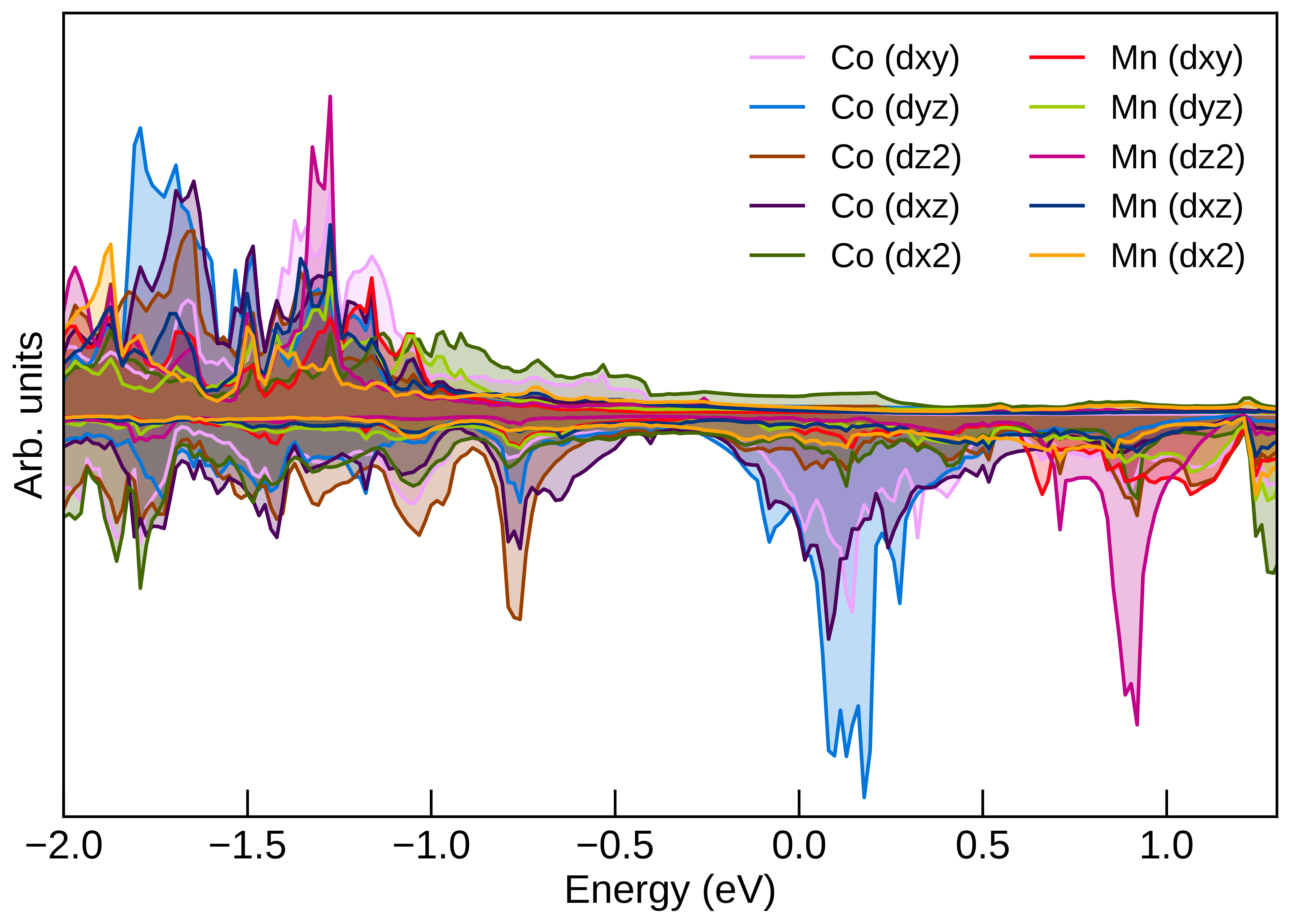}
\caption{Density of states of Co$_3$Mn$_2$Sn (MAE = -0.42 MJ/m$^3$). } \label{DOS-Sn}
\end{subfigure}

\caption{(Color online) Density of states of Co$_3$Mn$_2$X, with X = Al, Si, P, Ga, Ge, As, In, Sn, Sb}
 \label{fig:DOS-all}
\end{figure}

\begin{figure}[h!]

 \centering
 \includegraphics[scale=0.3]{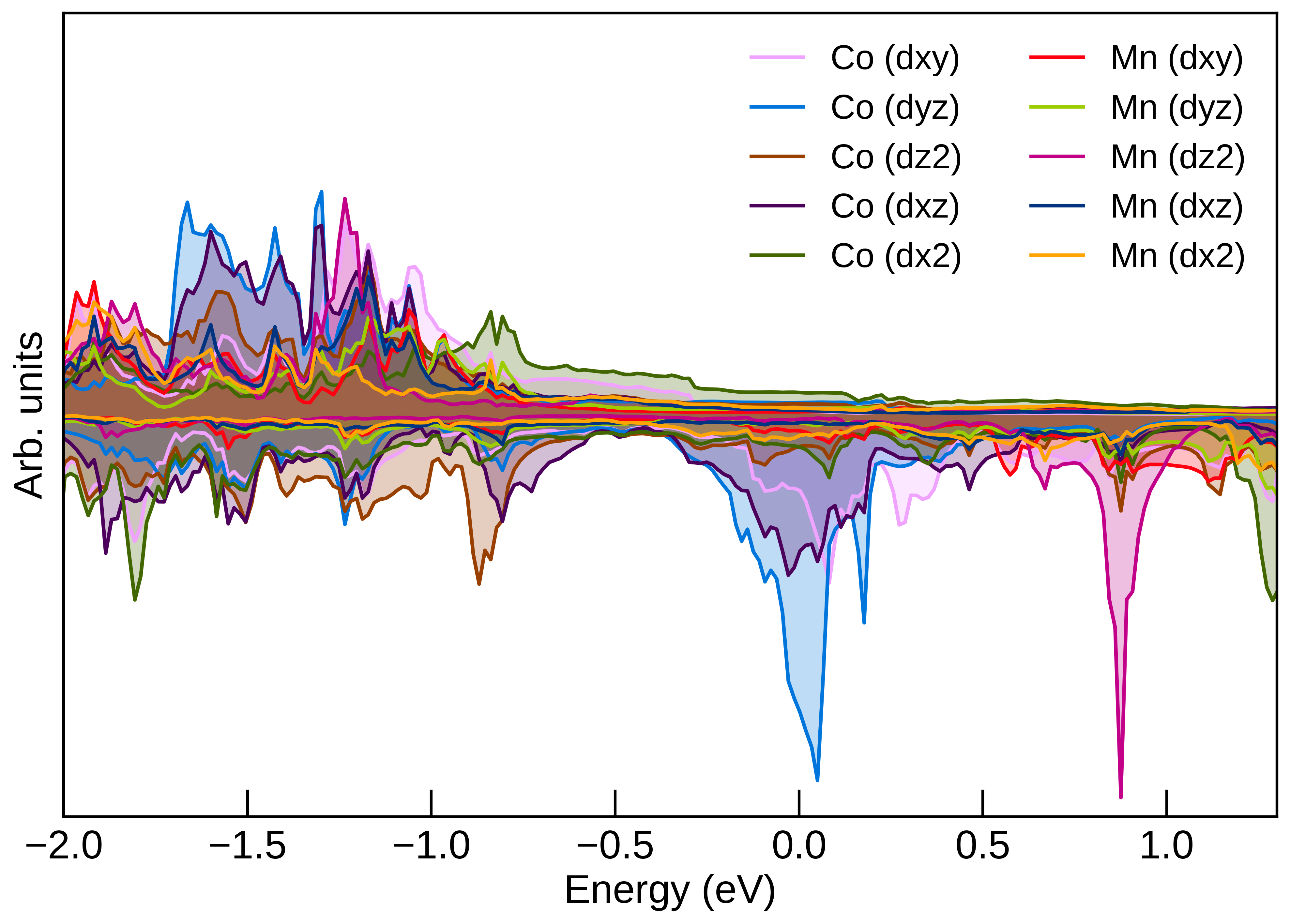}
 \caption{(Color online) Density of states of Co$_3$Mn$_2$Sb (MAE = -0.81 MJ/m$^3$).}
 \label{DOS-Sb}
\end{figure}

To understand the origin of the difference in MAE for Co$_3$Mn$_2$X (X = Al, Si, P, Ga, As, In, Sn, Sb, Tl, and Pb), it is electron states close to the Fermi level ($\epsilon_F$) one should focus on, since spin-orbit interaction that couples states just below $\epsilon_F$ to states just above $\epsilon_F$, are particularly important in deciding the magnetic anisotropy \cite{mat9}. To undertake this analysis we inspected partial densities of states (pDOS) around $\epsilon_F$. Figure~\ref{DOS_Ge} shows the pDOS for Co$_3$Mn$_2$Ge, which has a large uniaxial anisotropy of 1.44 MJ/m$^3$, and Fig.~\ref{DOS_As} shows the pDOS for Co$_3$Mn$_2$As, that has a large easy-plane anisotropy of -1.2 MJ/m$^3$. The pDOS curves for the other Co$_3$Mn$_2$X materials can be found in {\it Appendix~\ref{sec:dos}}, Figs.~\ref{fig:DOS-all}-\ref{DOS-Sb}. The majority spin channel of the d-states is essentially fully occupied for all the materials considered in this work which shows its inertness for the magnetic anisotropy. Instead significant contributions are expected for the minority spin channel, that has states on either side of $\epsilon_F$. In this spin channel there are large peaks corresponding to the  $d_{yz}$, $d_{xz}$, and $d_{xy}$ states of Co, close to the Fermi energy. These peaks are mostly empty for X = Al, Ga, Ge, In, and Sn, while they lie directly on $\epsilon_F$ for X = Si and Sb, and are mostly occupied for X = P, As. All the materials with uniaxial magnetic anisotropy (X = Al, Ga, Ge, In) have these large peaks of the DOS occupied, and we conclude that from a microscopic point of view, these electrons are decisive for the magnetic anisotropy. 

To analyze the MAE further we consider the difference in SOC energies with spin orientation along the {\it z} and {\it x} axes, $\Delta E_{\textrm{so}}=E_{\textrm{so}}^z-E_{\textrm{so}}^x$, separately for all Co and Mn atoms, see Table \ref{table:CoMnGe} (negative sign marks a contribution to uniaxial magnetic anisotropy). As expected, for most of {\it X}, $\Delta E_{\textrm{so}}$(Mn) is considerably smaller than that of Co atoms. The latter can be divided into two groups (noted by the subscripts in Table~\ref{table:CoMnGe}); the contribution to  $\Delta E_{\textrm{so}}$ from the two Co atoms positioned at (0.667, 0.833, 0.75) and (0.333, 0.166, 0.25) 6{\it h} (we denote them {\it type 2}) is different to that of the remaining four Co atoms ({\it type 1}). All Co atoms contribute to the uniaxial magnetic anisotropy for X = Al, Ga, and In. The values of $\Delta E_{\textrm{so}}$ are extremely low for both Co and Mn atoms in Co$_3$Mn$_2$P, which results in the negligibly small value of MAE. In the case of Co$_3$Mn$_2$Si and Co$_3$Mn$_2$As, all Co atoms give rise to the large values corresponding to easy-plane magnetocrystalline  anisotropy. For the remaining materials, the two groups of Co atoms have opposite signs of $\Delta E_{\textrm{so}}$.

We can also determine the d-orbitals that give the largest change to the SOC matrix element $\bra{\mu\sigma} \widehat{H}_{\textrm{so}} \ket{\mu'\sigma'}$ as the magnetization direction changes from  {\it z} to  {\it x}. Co$_3$Mn$_2$X with X = Al, Ga, Ge, and In, that exhibit uniaxial anisotropy, all present a similar picture, see Fig.~\ref{Ge_Co} ({\it Appendix~\ref{sec:soc}}) for Co$_3$Mn$_2$Ge (the rest of the figures can be found in {\it Appendix~\ref{sec:soc}}, Fig.~\ref{Al_Co}-\ref{In_Co}). The main contribution to the uniaxial anisotropy comes from $\bra{d_{x^2+y^2}} \widehat{H}_{\textrm{so}} \ket{d_{xy}}$ for all the Co atoms, even though there are additional contributions which are different for the Co atoms of {\it type 1} (Fig.~\ref{Ge_Co}, top) and {\it type 2} (Fig.~\ref{Ge_Co}, bottom). As expected (Table \ref{table:5}), for these transitions  to contribute to uniaxial anisotropy the states must be within the same spin channel. It is more difficult to distinguish any specific transitions for materials with the easy-plane anisotropy, see Fig.~\ref{As_Co} ({\it Appendix~\ref{sec:soc}}) for Co$_3$Mn$_2$As (the other materials can be found in {\it Appendix~\ref{sec:soc}},
Fig.~\ref{Al_Co}-\ref{In_Co}) -  there are several matrix elements promoting the negative sign in MAE. The most significant are $\bra{d_{x^2+y^2}} \widehat{H}_{\textrm{so}} \ket{d_{yz}}$, $\bra{d_{xz}} \widehat{H}_{\textrm{so}} \ket{d_{xy}}$, and $\bra{d_{z^2}} \widehat{H}_{\textrm{so}} \ket{d_{yz}}$ and their size varies depending on X.

\section{d-orbitals contribution to MAE}\label{sec:xyz}

Matrix elements $\bra{\mu\sigma}L\cdot S \ket{\mu'\sigma'}$ \cite{Abate,Takayama} determine the preferable direction of spin quantization axis.
Table \ref{table:5} lists the transitions between the states below and above $\epsilon_F$ which favour either z-axis (uniaxial) or xy-plane (planar) magnetic anisotropy along with the relative weight of each transition.

\begin{table*}[h!]
\centering
\caption{Transitions between the d-orbitals below and above $\epsilon_F$ which contribute to either  z-axis (uniaxial) or xy-plane (planar) magnetic anisotropy along with the relative weight of each transition, $\omega$}
\begin{tabular}{c|cc|cc} 
 \hline \hline
 Contributes & Same spin & $\omega$ & Opposite spin & $\omega$ \\ 
  & transition &  & transition &  \\
 \hline
Uniaxial & $d_{xy} \xrightarrow{}d_{x^2-y^2}$ & 1 & $d_{xy} \xrightarrow{} d_{xz}$, $d_{xy} \xrightarrow{} d_{yz}$ & 0.25  \\ 
anisotropy & $d_{yz} \xrightarrow{}d_{xz}$ & 0.25 &  $d_{xz} \xrightarrow{} d_{z^2}$, $d_{yz} \xrightarrow{} d_{z^2}$  & 0.75 \\
 &  &  & $d_{xz} \xrightarrow{}d_{x^2-y^2}$, $d_{yz} \xrightarrow{}d_{x^2-y^2}$ & 0.25 \\
\hline
xy-plane & $d_{xy} \xrightarrow{} d_{xz}$, $d_{xy} \xrightarrow{} d_{yz}$ & 0.25 & $d_{xy} \xrightarrow{}d_{x^2-y^2}$ & 1 \\
anisotropy &  $d_{xz} \xrightarrow{} d_{z^2}$, $d_{yz} \xrightarrow{} d_{z^2}$ & 0.75 & $d_{yz} \xrightarrow{}d_{xz}$ & 0.25 \\
 & $d_{xz} \xrightarrow{}d_{x^2-y^2}$, $d_{yz} \xrightarrow{}d_{x^2-y^2}$ & 0.25 &  &  \\
 \hline \hline
\end{tabular}
\label{table:5}
\end{table*}

\newpage
\section{DOS for Co$_3$Mn$_2$X (X = Al, Si, P, Ga, In, Sn, Sb)}\label{sec:dos}

Here we provide the DOS for the crystal structures Co$_3$Mn$_2$X (X = Al, Si, P, Ga, In, Sn, Sb, Tl, and Pb). These were calculated by replacing Ge in Co$_3$Mn$_2$Ge structure by the neighboring elements and relaxing their crystal structures. DOS of Co$_3$Mn$_2$Ge and Co$_3$Mn$_2$As are given above in the text. 

\newpage
\section{Change of the SOC matrix elements of Co when magnetization changes direction from {\it z} to  {\it x}  for Co$_3$Mn$_2$X (X = Al, Si, P, Ga, In)}\label{sec:soc}

\begin{figure*}[hbt!]
     \centering
     \begin{subfigure}[b]{0.38\textwidth}
         \centering
         \includegraphics[width=\textwidth]{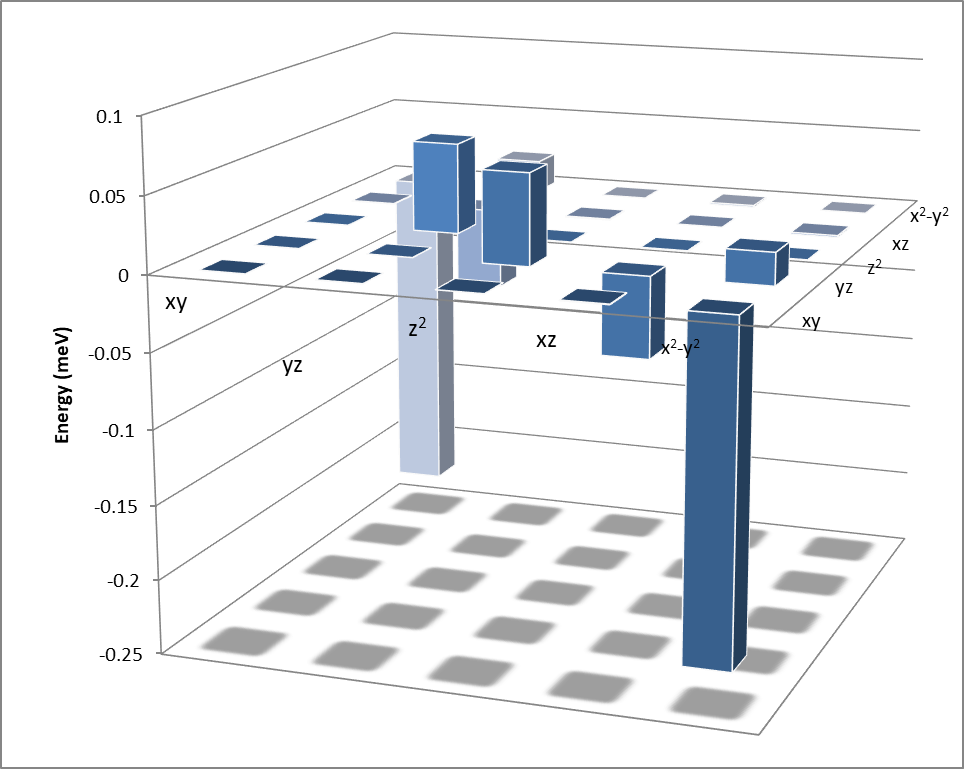}
         \label{Al_Co1}
     \end{subfigure}
     \hfill
     \begin{subfigure}[b]{0.38\textwidth}
         \centering
         \includegraphics[width=\textwidth]{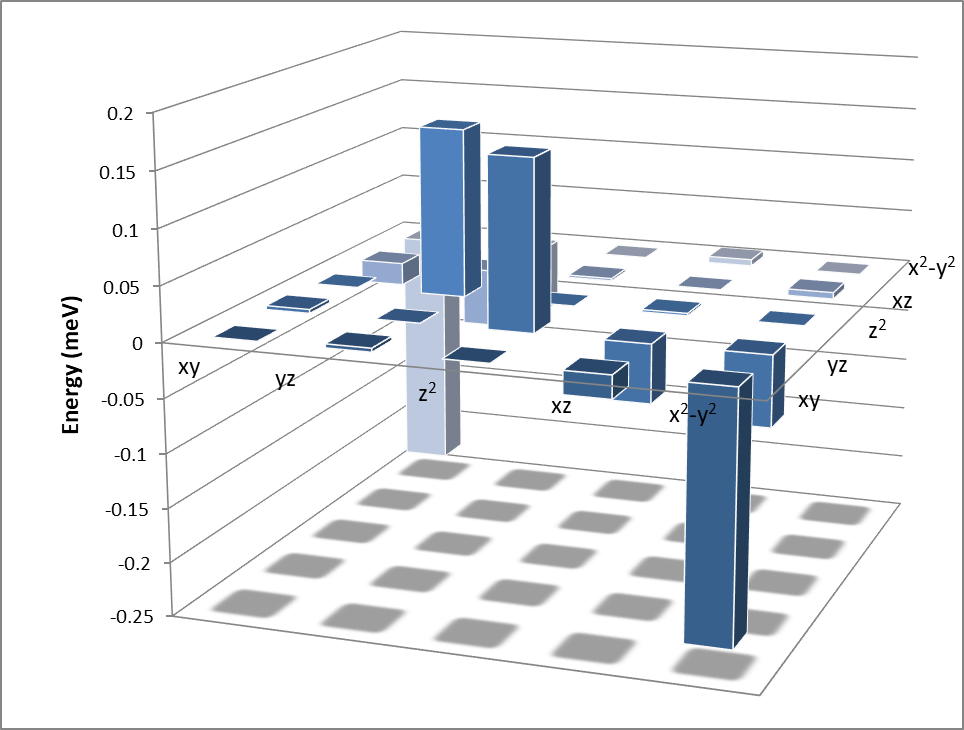}
         \label{Al_Co2}
     \end{subfigure}
        \caption{(Color online) Change of spin-orbit coupling matrix elements of Co, {\it type 1} (left), and Co, {\it type 2} (right) in Co$_3$Mn$_2$Al when magnetization changes direction from {\it z} to  {\it x}.}
        \label{Al_Co}
\end{figure*}

\begin{figure*}[hbt!]
     \centering
     \begin{subfigure}[b]{0.38\textwidth}
         \centering
         \includegraphics[width=\textwidth]{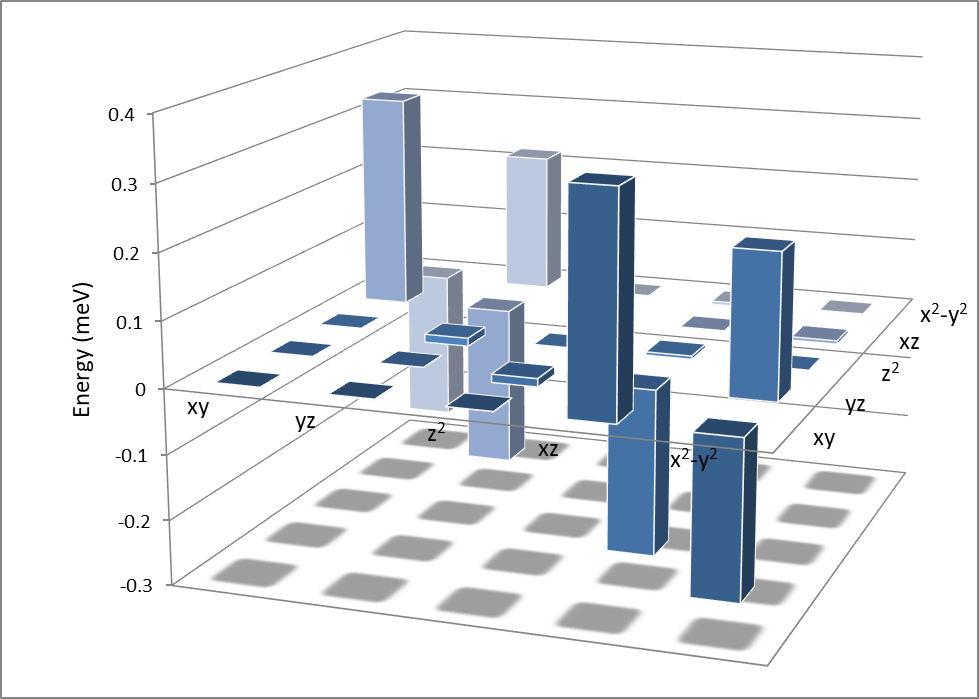}
         \label{Si_Co1}
     \end{subfigure}
     \hfill
     \begin{subfigure}[b]{0.38\textwidth}
         \centering
         \includegraphics[width=\textwidth]{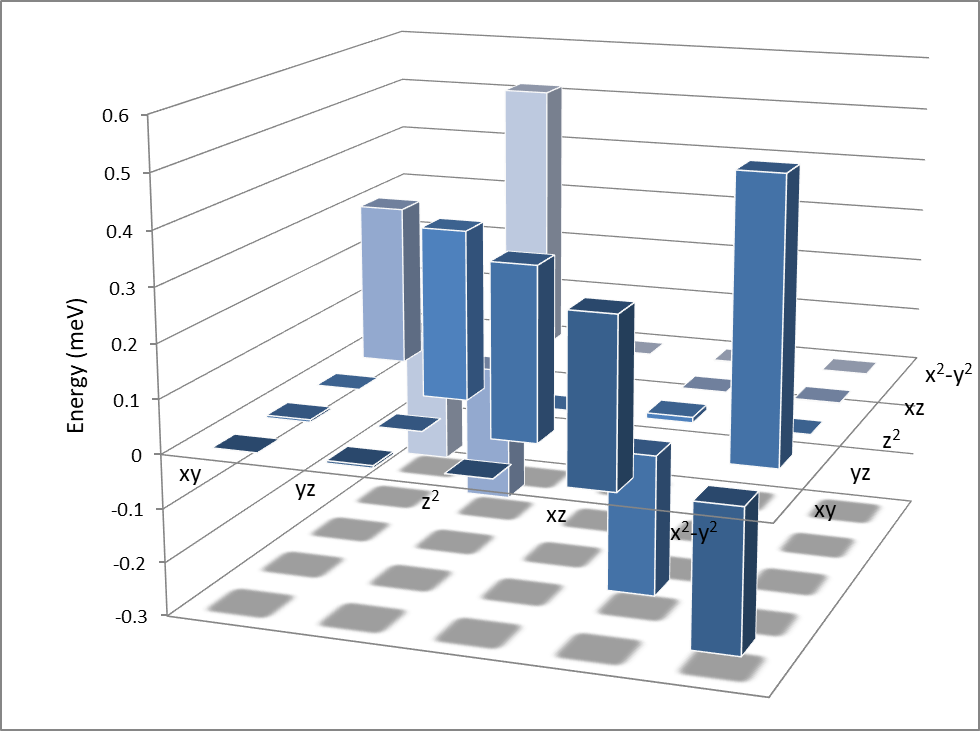}
         \label{Si_Co2}
     \end{subfigure}
        \caption{(Color online) Change of spin-orbit coupling matrix elements of Co, {\it type 1} (left), and Co, {\it type 2} (right) in Co$_3$Mn$_2$Si when magnetization changes direction from {\it z} to  {\it x}.}
        \label{Si_Co}
\end{figure*}

\begin{figure*}[hbt!]
     \centering
     \begin{subfigure}[b]{0.38\textwidth}
         \centering
         \includegraphics[width=\textwidth]{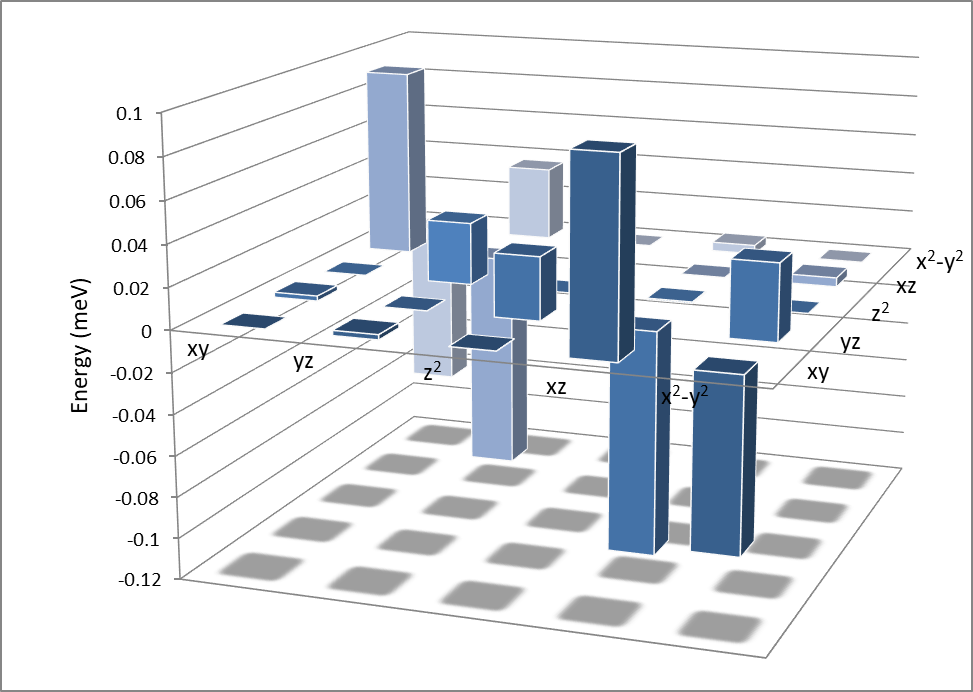}
         \label{P_Co1}
     \end{subfigure}
     \hfill
     \begin{subfigure}[b]{0.38\textwidth}
         \centering
         \includegraphics[width=\textwidth]{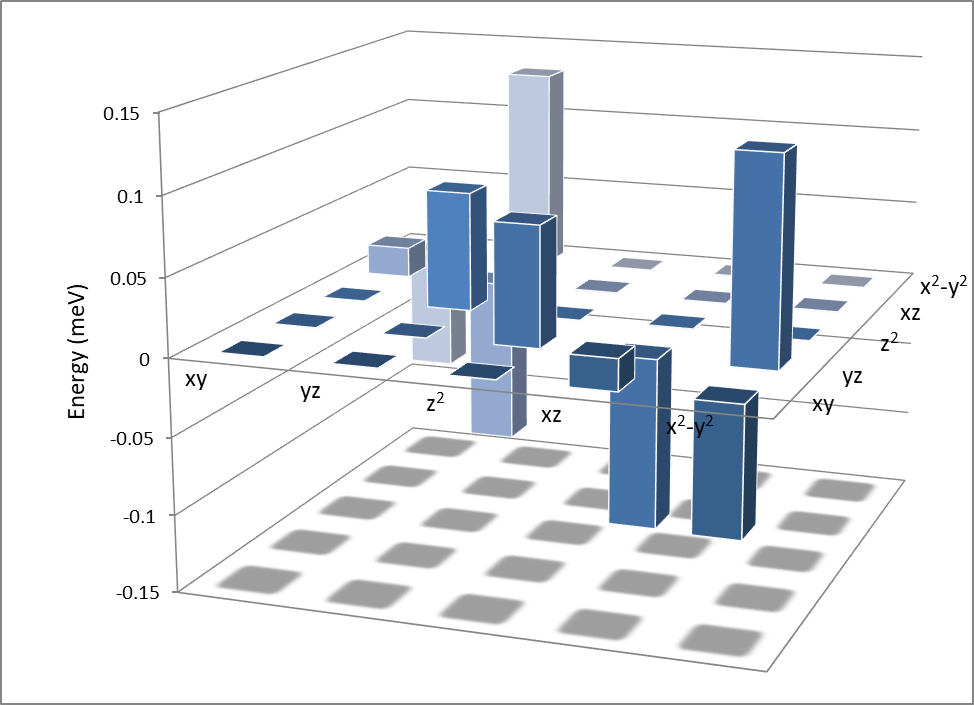}
         \label{P_Co2}
     \end{subfigure}
        \caption{(Color online) Change of spin-orbit coupling matrix elements of Co, {\it type 1} (left), and Co, {\it type 2} (right) in Co$_3$Mn$_2$P when magnetization changes direction from {\it z} to  {\it x}.}
        \label{P_Co}
\end{figure*}

\begin{figure*}[hbt!]
     \centering
     \begin{subfigure}[b]{0.38\textwidth}
         \centering
         \includegraphics[width=\textwidth]{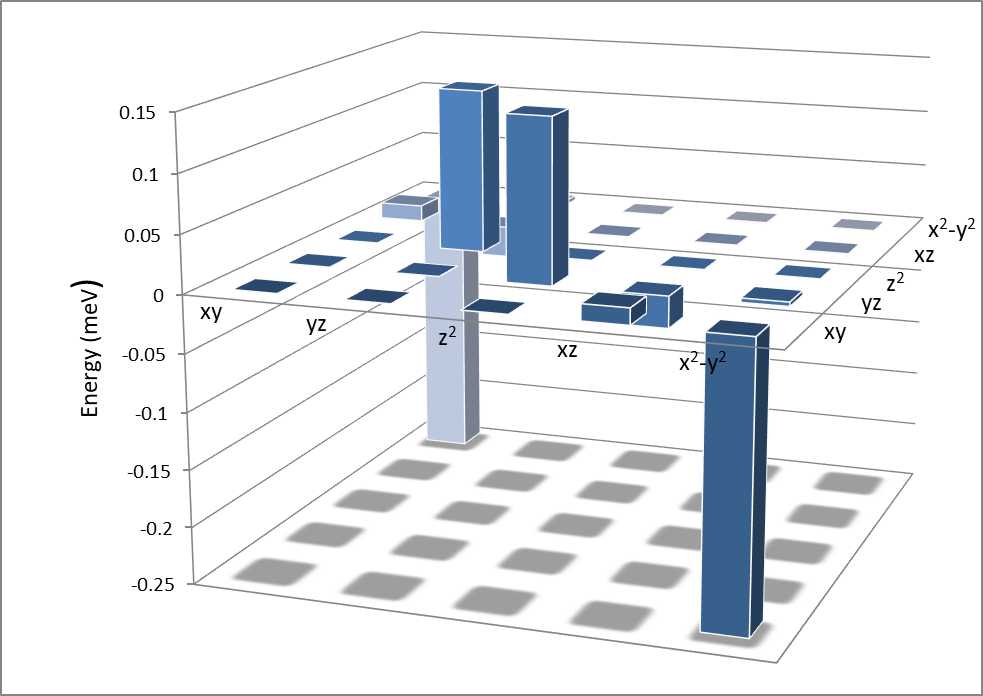}
         \label{Ga_Co1}
     \end{subfigure}
     \hfill
     \begin{subfigure}[b]{0.38\textwidth}
         \centering
         \includegraphics[width=\textwidth]{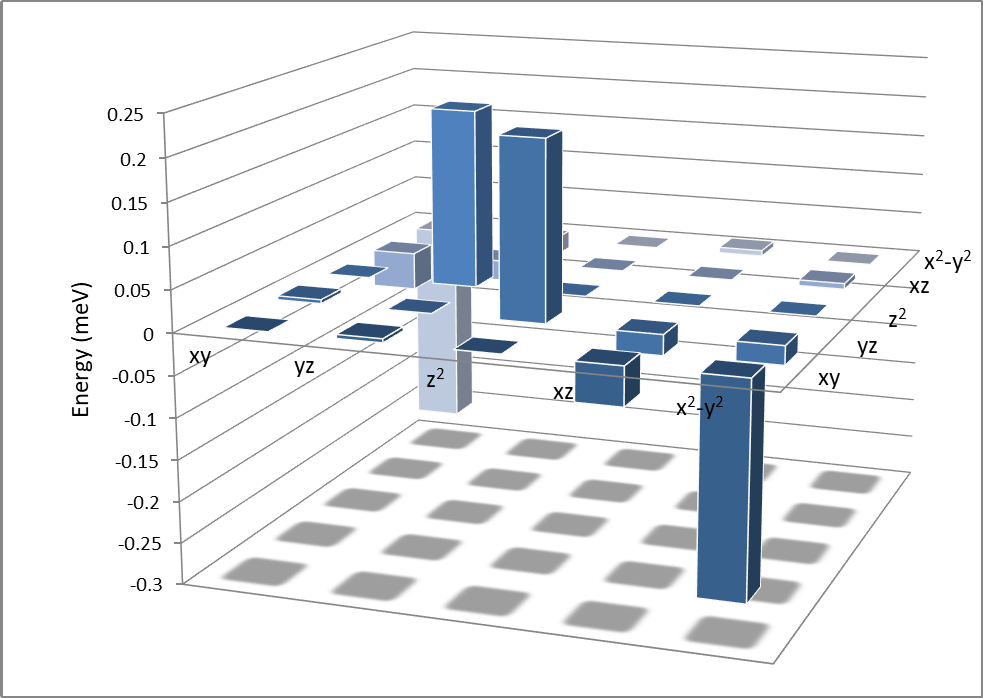}
         \label{Ga_Co2}
     \end{subfigure}
        \caption{(Color online) Change of spin-orbit coupling matrix elements of Co, {\it type 1} (left), and Co, {\it type 2} (right) in Co$_3$Mn$_2$Ga when magnetization changes direction from {\it z} to  {\it x}.}
        \label{Ga_Co}
\end{figure*}

\begin{figure*}[hbt!]
     \centering
     \begin{subfigure}[b]{0.38\textwidth}
         \centering
         \includegraphics[width=\textwidth]{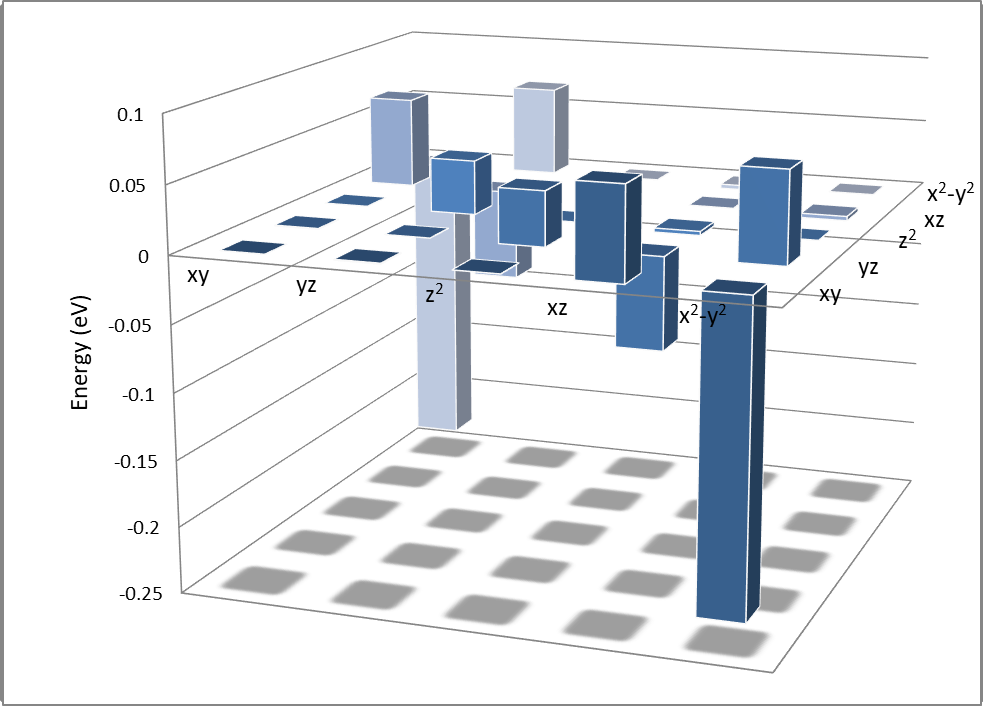}
         \label{Ge_Co1}
     \end{subfigure}
     \hfill
     \begin{subfigure}[b]{0.38\textwidth}
         \centering
         \includegraphics[width=\textwidth]{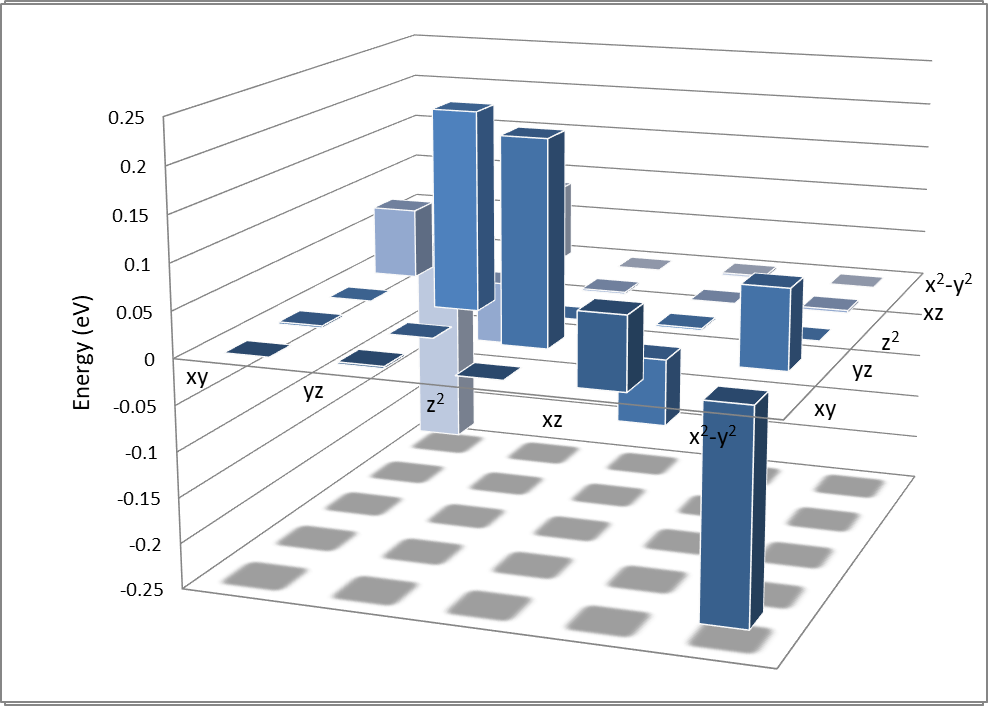}
         \label{Ge_Co2}
     \end{subfigure}
        \caption{(Color online) Change of spin-orbit coupling matrix elements of Co, {\it type 1} (top), and Co, {\it type 2} (bottom) in Co$_3$Mn$_2$Ge when magnetization changes direction from {\it z} to  {\it x}.}
        \label{Ge_Co}
\end{figure*}

\begin{figure*}[hbt!]
     \centering
     \begin{subfigure}[b]{0.38\textwidth}
         \centering
         \includegraphics[width=\textwidth]{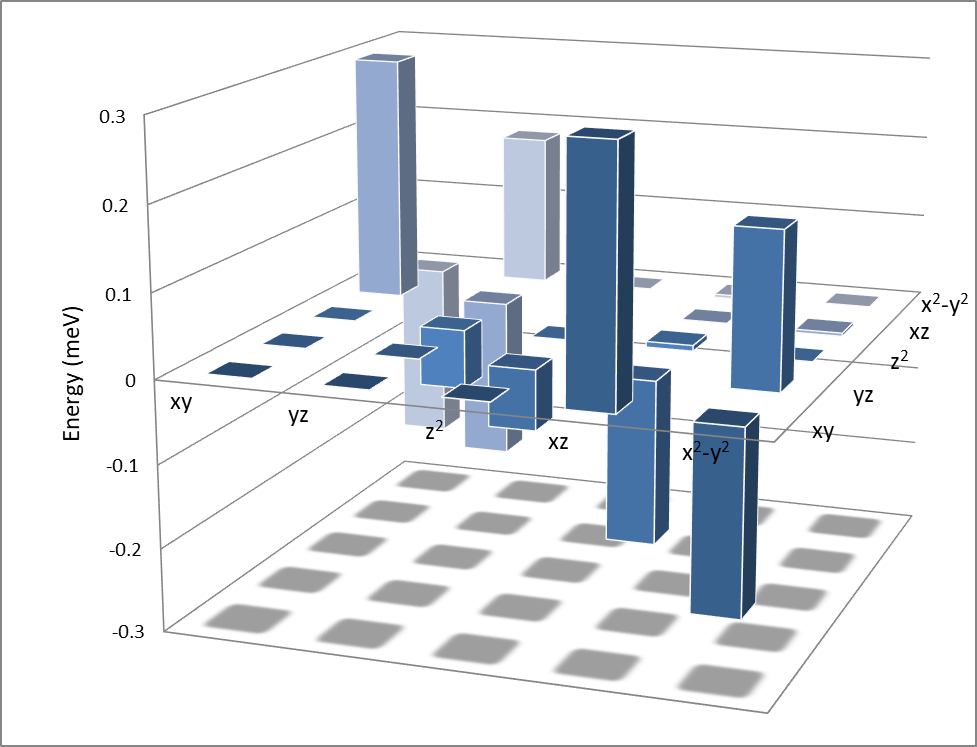}
         \label{As_Co1}
     \end{subfigure}
     \hfill
     \begin{subfigure}[b]{0.38\textwidth}
         \centering
         \includegraphics[width=\textwidth]{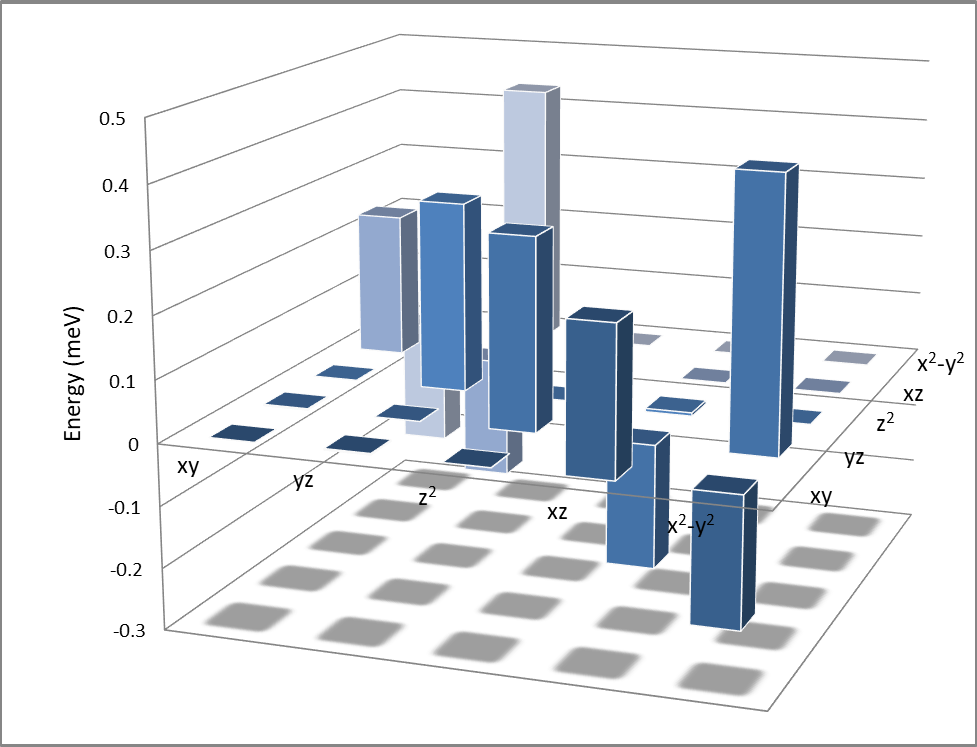}
         \label{As_Co2}
     \end{subfigure}
        \caption{(Color online) Change of spin-orbit coupling matrix elements of Co, {\it type 1} (top), and Co, {\it type 2} (bottom) in Co$_3$Mn$_2$As when magnetization changes direction from {\it z} to  {\it x}.}
        \label{As_Co}
\end{figure*}

\begin{figure*}[hbt!]
     \centering
     \begin{subfigure}[b]{0.38\textwidth}
         \centering
         \includegraphics[width=\textwidth]{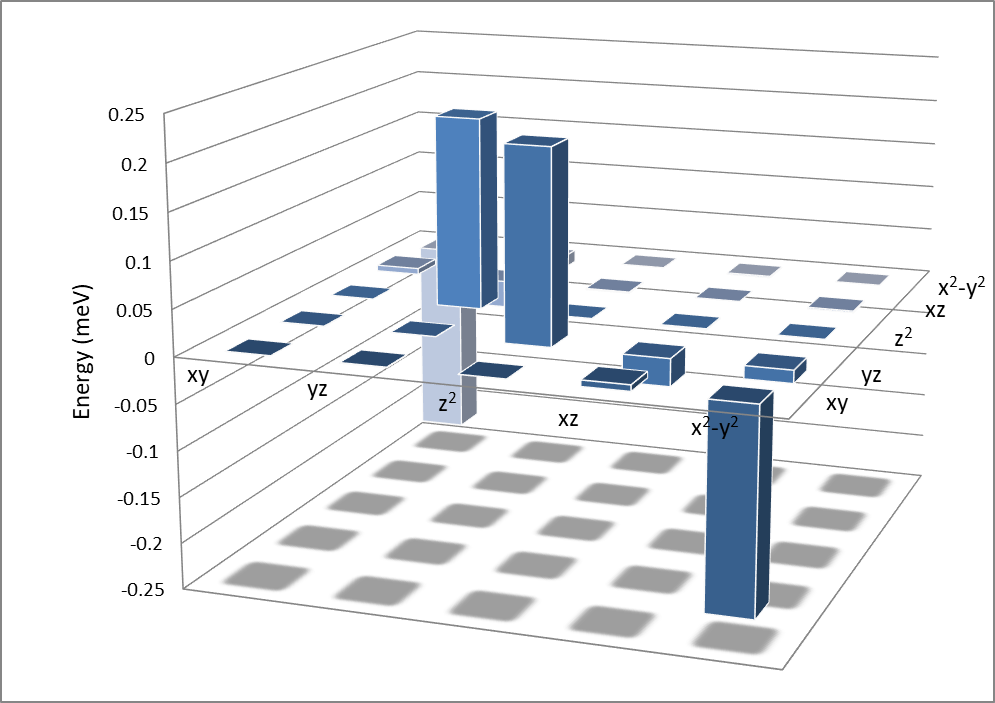}
         \label{In_Co1}
     \end{subfigure}
     \hfill
     \begin{subfigure}[b]{0.38\textwidth}
         \centering
         \includegraphics[width=\textwidth]{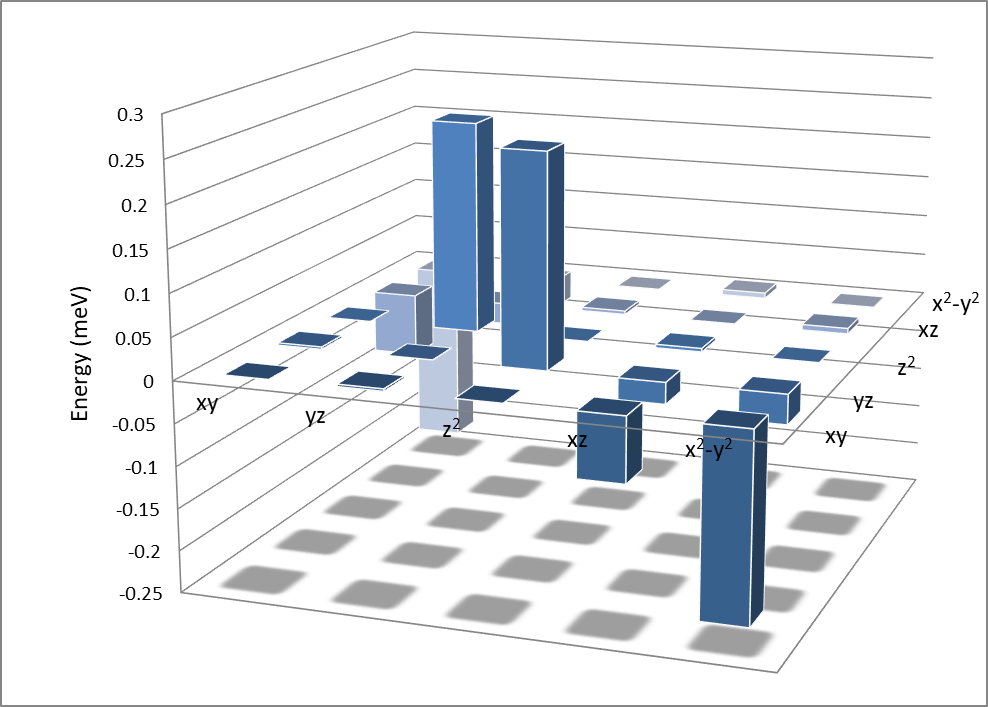}
         \label{In_Co2}
     \end{subfigure}
        \caption{(Color online) Change of spin-orbit coupling matrix elements of Co, {\it type 1} (left), and Co, {\it type 2} (right) in Co$_3$Mn$_2$In when magnetization changes direction from {\it z} to  {\it x}.}
        \label{In_Co}
\end{figure*}

To determine the main orbital contribution to MAE we calculate the change in SOC matrix element $\bra{\mu\sigma} \widehat{H}_{\textrm{so}} \ket{\mu'\sigma'}$ for transitions between different {\it d}-orbitals as magnetization direction changes from {\it z} to  {\it x}. Fig.~\ref{Al_Co}-\ref{In_Co} show contributions from {\it type 1} Co atoms (top) and {\it type 2} Co atoms (bottom).

\section{The properties of the previously reported phases for Co-Mn-Ge system}\label{sec:comnge_previous}

The previously reported crystallographic phases for the Co--Mn--Ge system are presented in Table \ref{table:phases} together with their crystal structure and some of the magnetic properties. 
We will briefly describe some of the key features of the structural and magnetic properties of Co--Mn--Ge systems reported in the literature, which are expected to be relevant to the Co$_3$Mn$_2$Ge. 
It should be noted that this outline is \textbf{not} intended as a review, but rather as a summary of the known properties of Co--Mn--Ge phases, which we find relevant to understanding the properties of Co$_3$Mn$_2$Ge.

 \begin{table*}[t]
\centering
\caption{Co--Mn--Ge phases reported previously, with their crystal structures and transition temperatures.}
\begin{tabular}{l l l} 
 \hline \hline
Compound & Entry prototype,  & Remarks  \\ 
 & SGR Symbol and number & and comments \\
 \hline
CoMnGe & TiNiSi, {\it Pnma} (62) & Low-temperature phase, stable below 650 K \cite{KANOMATA1995131}; \\
 &  & $T_{\rm C}$ = 337 \cite{KAPRZYK1990267}.   \\
CoMnGe & BeZrSi, \textit{P}6$_3$/\textit{mmc} (194) & High-temperature phase, stable above 650 K  \cite{KANOMATA1995131}; \\
 & & collinear \cite{Szytula1981} with $T_{\rm C}$ = 283 K \cite{KAPRZYK1990267}, 334 K \cite{KANOMATA1995131}. \\
 & & Transformation with $T_{\rm M}$=$T_{\rm C}$ leads to large/giant magnetocaloric \\
 & & effect \cite{Trung2010,Samanta2012,MA2012135}. \\
Co$_x$Mn$_{1-x}$Ge & Ni$_2$In, \textit{P}6$_3$/\textit{mmc}(194) & Spin reorientation depending on the amount of Co; can be easy axis,\\
 & & easy plane, as well as hard/easy equally in all directions \cite{Markin2008}. \\
Co$_2$MnGe & Cu$_2$MnAl, {\it Fm$\bar{3}$m} (225) & The increase in quenching temperature leads to increase in Mn-Ge  \\
 & & disorder \cite{Kogachi2009}. Disorder causes the decrease in the magnetization \cite{Kogachi2009};\\
 & & higher quenching temperature rates produce lower magnetization \cite{Kogachi2009}. \\
 & & High Mn-Ge disorder, thin films \cite{Collins2015}, enrichment of 5 at \% of Ge in \\
 & & Co$_{0.5}$Mn$_{0.25}$Ge$_{0.25}$ gives the highest degree of chemical ordering \cite{Collins2015}. \\
 & & Chemical disorder in Co$_2$MnGe decreases $T_{\rm C}$  \cite{Okubo2010}. \\
 & & Strong ordering in Co$_2$MnGe  \cite{Webster1971}, almost all magnetic moment comes \\
 & & from Mn (3.58 $\mu_B$ vs. 0.75 $\mu_B$ for Co). \\
Co$_x$Ni$_{1-x}$MnGe & TiNiSi, {\it Pnma} (62) & Samples can be either AFM or FM or non-collinear FM depending on \\
 & & $x$ and the annealing temperature \cite{Nizio1981}. \\
Co$_3$Mn$_2$Ge & Mg$_2$Cu$_3$Si, \textit{P}6$_3$/\textit{mmc} (194) & Ordered crystallographic model \cite{CoMnGe} \\
Co$_3$Mn$_2$Ge & MgZn$_2$, \textit{P}6$_3$/\textit{mmc} (194) & Disordered crystallographic model \cite{CoMnGe} \\
Co$_3$Ge$_5$Mn$_9$ & Mn$_9$Co$_3$Ge$_5$, \textit{R}32{\it h}, (155) & No magnetic properties reported. \cite{VENTURINI2014886} \\
 \hline \hline
\end{tabular}
\label{table:phases}
\end{table*}

CoMnGe exists in two stable phases; the low-temperature orthorhombic CoMnGe (TiNiSi type) is stable below the martensitic temperature ($T_{\rm M}$) of 650 K  \cite{KANOMATA1995131} while the high-temperature hexagonal CoMnGe (BeZrSi type) is stable above it \cite{KANOMATA1995131}. 
The reported $T_{\rm C}$-values of the high- and low-temperature phases vary slightly, for instance, Kaprzyk and Niziol  \cite{KAPRZYK1990267} report $T_{\rm C}$ = 337 K for CoMnGe (TiNiSi) and $T_{\rm C}$ = 287 K for CoMnGe (BeZrSi).
As it is possible to tune $T_{\rm M}$ to coincide with $T_{\rm C}$  \cite{KANOMATA1995131,Wang2006,FANG2007453},  a large/giant magnetocaloric effect can be achieved at the room temperature \cite{Trung2010,Samanta2012,MA2012135}. 

CoGeMn is reported to exhibit a collinear  \cite{Szytula1981} magnetic state.
However, depending on the Co:Mn ratio, Co$_x$Mn$_{1-x}$Ge samples can possess not only an easy-axis or easy-plane magnetic anisotropy but also an intermediate state  \cite{Markin2008}, i.e.  spin reorients with Co:Mn ratio.
The substitution of Ni (Co$_x$Ni$_{1-x}$MnGe) produces the even more diverse results; the samples can be AFM, collinear FM, and  non-collinear FM, depending on the amount of Ni and the annealing temperature \cite{Nizio1981}.

Co$_2$MnGe is a full Heusler compound which exhibits some common features with the prototypical full Heusler Cu$_2$MnAl.
These include a high $T_{\rm C}$ and certain peculiarities related to its structural and magnetic properties derived from the range of chemical environments, disorder, vacancies, and stacking faults.
Using anomalous X-ray diffraction techniques Collins \textit{et al.} \cite{Collins2015} were able to demonstrate that thin films of Co$_2$MnGe grown on Ge (111) show the preference of Mn--Ge site swapping along with the small ($< 0.5$ \%) site-interchange of Co--Mn.
The authors were also able to show that over--stoichiometry of Ge leads to the decrease in chemical disorder, vacancies, and stacking faults.
It is important to note, that they were not able to grow the Co$_3$Mn$_2$Ge hexagonal phases due to a too large lattice mismatch with the Ge (111) surface.

Kogachi \textit{et al.}  \cite{Kogachi2009} showed that the increase in sample quenching temperature gives rise to Mn--Ge disorder and results in a lower magnetization at 4.2 K.
Okubo \textit{et al.}  \cite{Okubo2010} were able to demonstrate that the aforementioned disorder brings on a considerable decrease in $T_{\rm C}$. Webster  \cite{Webster1971}, on the other hand, had earlier reported that there was no sign of chemical disorder in Co$_2$MnGe and that magnetic moment originated primarily from Mn (3.58 $\mu_B$); Co only contributes 0.75 $\mu_B$ only.
Some of these ''discrepancies'' reported previously are thus likely due to the difference in the preparation of the samples, but they also show that Co$_2$MnGe exists in a number of states ranging from a collinear ferromagnet with a high $T_{\rm C}$ to a virtually non-magnetic material, depending on the chemical disorder. 

Lastly, Co$_3$Ge$_5$Mn$_9$ is reported to be a rhombohedral phase with no magnetic properties reported \cite{VENTURINI2014886}.

\end{document}